\documentclass[manuscript]{acmart} 

\usepackage[skins]{tcolorbox}
\usepackage{graphicx}
\usepackage{listings}
\usepackage{url}
\usepackage{booktabs} 
\usepackage{multirow}
\usepackage{dcolumn}
\usepackage{color}
\usepackage{colortbl}
\usepackage{pifont}
\usepackage{hyphenat}
\usepackage{balance}
\usepackage{subfig}

\usepackage{tcolorbox,fancyvrb,xcolor,tikz}
\usepackage[linesnumbered, ruled, vlined]{algorithm2e}
\usepackage{amsmath,graphicx}

\newcommand*{\normal}{\mathnormal{N}}
\newcommand*{\uniform}{\mathnormal{Unif}}

\usepackage{nicefrac}

\usepackage{xspace}
\newcommand{\sysname}{ETAP\xspace}

\usepackage{tabularx,ragged2e}
\newcolumntype{C}[1]{>{\centering}m{#1}}
\newcolumntype{g}{>{\columncolor{babyblueeyes!25}}c}
\newcolumntype{Y}{>{\centering\arraybackslash}X}
\newcolumntype{R}{>{\raggedleft\arraybackslash}X}

\definecolor{darkgreen}{rgb}{0.0, 0.5, 0.0}
\definecolor{babyblueeyes}{rgb}{0.63, 0.79, 0.95}
\newcommand{\cmark}{\textcolor{darkgreen}{\ding{51}}}

\newcommand{\xmark}{\textcolor{red}{\ding{55}}}

\long\def\com#1{}

\AtBeginDocument{%
  \providecommand\BibTeX{{%
    \normalfont B\kern-0.5em{\scshape i\kern-0.25em b}\kern-0.8em\TeX}}}

\setcopyright{acmcopyright}
\copyrightyear{2022}
\acmYear{2022}
\acmDOI{XXXXXXX.XXXXXXX}

\begin{document}

\title{ETAP: Energy-aware Timing Analysis of Intermittent Programs}

\author{Ferhat Erata}
\affiliation{%
  \institution{Yale University}
  \city{Connecticut}
  \country{USA}}
\email{ferhat.erata@yale.edu}

\author{Arda Goknil}
\affiliation{%
  \institution{SINTEF Digital}
  \city{Oslo}
  \country{Norway}}
\email{arda.goknil@sintef.no}

\author{Eren Y{\i}ld{\i}z}
\affiliation{%
 \institution{Ege University}
 \city{Izmir}
 \country{Turkey}}
\email{eren.yildiz@ege.edu.tr}

\author{Kas{\i}m Sinan Y{\i}ld{\i}r{\i}m}
\affiliation{%
  \institution{University of Trento}
  \city{Trento}
  \country{Italy}}
\email{kasimsinan.yildirim@unitn.it}

\author{Ruzica Piskac}
\affiliation{%
  \institution{Yale University}
  \city{Connecticut}
  \country{USA}}
\email{ruzica.piskac@yale.edu}

\author{Jakub Szefer}
\affiliation{%
  \institution{Yale University}
  \city{Connecticut}
  \country{USA}}
\email{jakub.szefer@yale.edu}

\author{Gökçin Sezgin}
\affiliation{%
  \institution{UNIT Information Technologies R$\&$D Ltd.}
  \city{Izmir}
  \country{Turkey}}
\email{gokcin.sezgin@unitbilisim.com}

\begin{abstract}
Energy harvesting battery-free embedded devices rely only on ambient energy harvesting that enables stand-alone and sustainable IoT applications. 
These devices execute programs when the harvested ambient energy in their energy reservoir is sufficient to operate and stop execution abruptly (and start charging) otherwise. These intermittent programs have varying timing behavior under different energy conditions, hardware configurations, and program structures. This paper presents Energy-aware Timing Analysis of intermittent Programs (\sysname), a probabilistic symbolic execution approach 
that analyzes the timing and energy behavior of intermittent programs at compile time. 
\sysname symbolically executes the given program while taking time and energy cost models for ambient energy and dynamic energy consumption into account.
We evaluated \sysname on several intermittent programs and compared the compile-time analysis results with executions on real hardware. 
The results show that \sysname's normalized prediction accuracy is 
99.5\%, and it speeds up the timing analysis by at least two orders of magnitude compared to manual testing.
\end{abstract}

\ccsdesc[500]{Software and its engineering~Software maintenance tools}
\ccsdesc[500]{Computing methodologies~Model development and analysis}
\ccsdesc[500]{Software and its engineering~Software maintenance tools}
\ccsdesc[500]{Computing methodologies~Symbolic and algebraic algorithms}
\ccsdesc[500]{Computer systems organization~Embedded software}

\keywords{Intermittent computing, energy harvesting, symbolic execution, timing and energy estimation}

\maketitle

\section{Introduction}
\label{sec:introduction}

Advancements in energy-harvesting circuits and ultra-low-power microelectronics enable modern battery-free devices that operate by using only harvested energy. 
Recent works have demonstrated several promising applications of these devices, such as battery-free sensing~\cite{flicker, dnn, zygarde} and autonomous robotics~\cite{ink}. These devices also pave the way for new stand-alone, sustainable applications and systems, such as body implants~\cite{gutruf2018fully} and long-lived wearables~\cite{capband}, where continuous power is not available and changing batteries is difficult.

Battery-free devices can harvest energy from several sporadic and unreliable ambient sources, such as solar~\cite{camaroptera}, radio-frequency~\cite{flicker}, and even plants~\cite{konstantopoulos2015converting}. The harvested energy is accumulated in a tiny capacitor that can only store a marginal amount of energy. The capacitor powers the microcontroller, sensors, radio, and other peripherals. These components drain the capacitor frequently. When the capacitor drains out, the battery-free device experiences a power failure and switches to the harvesting of energy to charge itself. When the stored energy is above a threshold, the device reboots and becomes active again to compute, sense, and communicate. Successive charge-discharge cycles lead to frequent power failures, and in turn, \emph{intermittent execution} of programs. 
Each power failure leads to the loss of the device's volatile state, i.e., registers, memory, and peripheral properties. 

To countermeasure power failures and execute programs intermittently so that the computation can progress and memory consistency is preserved, researchers developed two recovery approaches supported by runtime environments. The first one is to store the volatile program state into non-volatile memory with checkpoints placed in the program source~\cite{mementos, quickrecall, hibernus++, ratchet, hicks2017clank, dino, harvos, chinchilla}. Another one is the task-based programming model~\cite{chain, ink, alpaca}, in which programs are a collection of idempotent and atomic tasks. 
Several studies extended these recovery approaches by considering different aspects, e.g., I/O support~\cite{coati}, task scheduling~\cite{catnap, zygarde}, adaptation~\cite{coala}, and timely execution~\cite{mayfly, tics}.

\begin{figure}
    \centering
    \noindent\includegraphics[width=0.75\columnwidth]{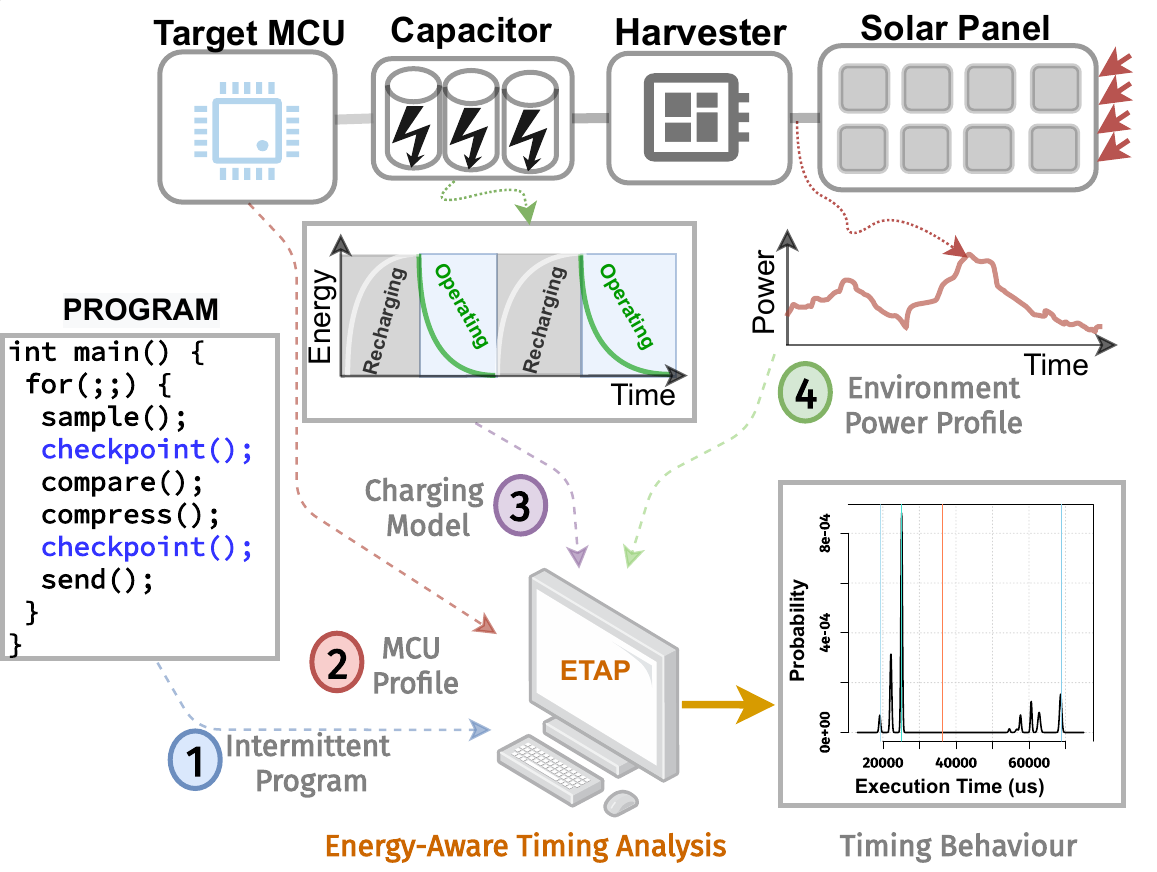} 
    \caption{The execution time of an intermittent program depends on energy-harvesting dynamics, capacitor size, the power consumption of the target microcontroller (MCU), and checkpoint overhead. Without hardware deployment, \sysname predicts timing behavior of an intermittent program.}
    \label{fig:intro}
\end{figure}

Considerable research has been devoted to compile-time analysis to find bugs and anomalies of intermittent programs~\cite{ibis, sceptic} and structure them (via effective task splitting and checkpoint placement) based on worst-case energy consumption analysis~\cite{cleancut, epic}.
Despite these efforts, 
no attention has been paid to 
analyzing the timing behavior of intermittent programs, affected by several factors such as the energy availability of the deployment environment, the power consumption of the target hardware, capacitor size, program input space, and program structure (checkpoint placement). Without such an analysis, programmers will never know \emph{at compile-time} if their intermittent programs execute as they are intended to do in a real-world deployment. 
Worse still, it is extremely costly and time-consuming to analyze the timing behavior of intermittent programs on real deployments because programmers need to run the programs multiple times on the target hardware with various ambient energy profiles, hardware configurations, and program inputs and structuring. 

As an example, consider a deployment environment with low ambient energy and frequent power failures.
The computing progresses slowly due to long charging periods, and in turn, the program throughput may not meet the expectations (e.g., a batteryless long-range remote visual sensing system takes a picture every 5 minutes and transmits the relevant ones every 20 minutes).  
If the program execution time is not what is expected, programmers might increase the capacitor size, change the target hardware, or remove some checkpoints. 
These changes may not always lead to what is intended (e.g., the bigger the capacitor size is, the longer the charging takes). 
Therefore, programmers might have to deploy their programs into the target platform several times and run them in the operating environment to check if they meet the desired throughput. 
Having estimations at a low cost \emph{without need to test and re-test multiple times on real hardware}, they can rapidly restructure the program (e.g., add or remove checkpoints) or reconfigure the hardware (e.g., change the microcontroller, its frequency, or capacitor size) to improve the program execution time and, in turn, the throughput.

\subsubsection*{Goal and Challenges.} 
Our goal in this paper is to estimate statically, \emph{at compile-time}, the timing behavior of intermittent programs considering different energy modes, hardware configurations, program input space, and program structure.
The state-of-the-art approaches exploit stochastic models to represent the dynamic energy harvesting environments, e.g., several models based on empirical data for RF~\cite{mishra2015charging, naderi2015wireless} and solar~\cite{preact} energy harvesting environments. Similarly, several techniques~\cite{epic, cleancut} extract the energy consumption characteristics of the target microcontroller instruction set to model its energy consumption. 
These two stochastic models, capacitor size, charging/discharging model, program input space, and program structure form a multi-dimensional space. The main challenge which we solve is to devise a program analysis solution that considers this multi-dimensional space to derive, \emph{at compile-time}, the execution time probabilities of the given intermittent program. 

Probabilistic symbolic execution~\cite{geldenhuys2012probabilistic} is an ideal solution to tackle this challenge since it is a compile-time program analysis solution quantifying the likelihood of properties of program states concerning program non-determinism. 
However, existing techniques consider only typical program non-determinism, e.g., program input probability distribution. Therefore, we need a dedicated symbolic execution technique that integrates intermittent program non-determinism (charging/discharging model and environment energy profile) and other intermittent program characteristics (capacitor size and program structure). 

\subsubsection*{Contributions.} In this paper, we propose, develop, and assess \textit{Energy-aware Timing Analysis of intermittent Programs (\sysname)}, a probabilistic symbolic execution approach for estimating the timing behavior of intermittent programs. We give an overview of \sysname's functionality in Figure~\ref{fig:intro}. \sysname generates the execution time probability distributions of each function in the input program 
Programmers may annotate programs with timing requirements for the execution time between any two lines of code or the time required to collect and process a given amount of data. \sysname also reports the execution time probability distributions for those requirements. It supports the checkpoint programming model but can easily be extended to analyze task-based programs. 
\sysname extends Clang~\cite{Clang} to recognize \sysname-specific configurations and uses LLVM compiler~\cite{lattner2004llvm}. It employs libraries of R environment~\cite{R} for symbolic computations and SMT~\cite{z3} solver to detect infeasible paths. Our symbolic execution technique is designed not to be specific to any microcontroller architecture.  

We compared our analysis to the program executions in a radio-frequency energy-harvesting testbed setup.
The evaluation results show that \sysname's normalized prediction accuracy is 99.5\%. To summarize, our contributions include: 
\begin{enumerate}

    \item \textbf{Novel Analysis Approach.} We introduce a novel probabilistic symbolic execution approach that predicts the execution times of intermittent programs and their timing behavior considering intermittent execution dynamics.
    
    \item \textbf{New Analysis Tool.} We introduce the first compile-time analysis tool that enables programmers to analyze the timing behavior of their intermittent programs without target platform deployment.

    \item \textbf{Evaluation on Real Hardware.} \sysname correctly predicts the execution time of intermittent programs with a maximum prediction error less than 1.5\% and speeds up the timing analysis by at least two orders of magnitude compared to manual testing.
    
\end{enumerate}
We share \sysname as a publicly available tool 
with the research community. We believe that \sysname is a significant attempt to provide the missing design-time analysis tool support for developing intermittent applications.

\section{Motivation and Background}
\label{sec:background}


A typical battery-free device such as WISP~\cite{wisp}, WISPCam~\cite{wispcam}, Engage~\cite{engage}, and Camaroptera~\cite{camaroptera} includes (i) an energy harvester converting incoming ambient energy into electric current, (ii) energy storage, typically a capacitor, to store the harvested energy to power electronics, and (iii) an ultra-low-power microcontroller orchestrating sensing, computation, and communication. The microcontrollers in these platforms, e.g., MSP430FR series~\cite{MSP430FR5994}, comprise a combination of volatile (e.g., SRAM) and non-volatile (e.g., FRAM~\cite{FRAM}) memory to store data that will persist upon power failures.

\subsection{Programming Intermittent Systems}
Battery-free platforms operate \emph{intermittently} due to frequent power failures. Several programming models (supported by runtimes) have been proposed to mitigate the effects of unpredictable power failures and enable \emph{computation progress} while preserving \emph{memory consistency} (e.g.,~\cite{chain,tics,ink}). Generally speaking, these models \emph{backup} the volatile state of the processor into the non-volatile memory so that the computation can be recovered from where it left upon reboot. Moreover, \emph{memory consistency} should also be ensured so that the backed-up state in the non-volatile memory will not be different from the volatile one, or vice versa. 

Programming models for intermittent computing have two classes. \emph{Checkpointing} systems~\cite{tics} snapshot the volatile state, i.e., the values of registers, stack, and global variables, in persistent memory at specific points---either defined by the programmer at compile-time or decided at runtime. 
Thanks to checkpoints, the state of the computation can be recovered after a power failure using the snapshot of the system state. \emph{Task-based} systems~\cite{chain,ink} employ a static task-based model. The programmer decomposes a program into a collection of tasks at compile-time and implements a task-based control flow (connecting task outputs with task inputs). The runtime keeps track of the active task, restarts it upon recovery from intermittent power failures, guarantees its atomic completion, and then switches to the next task in the control flow.

\begin{figure}
    \centering
    \includegraphics[trim={0.00cm 0.2cm 0.4cm 0.5cm},clip,width=0.75\columnwidth]{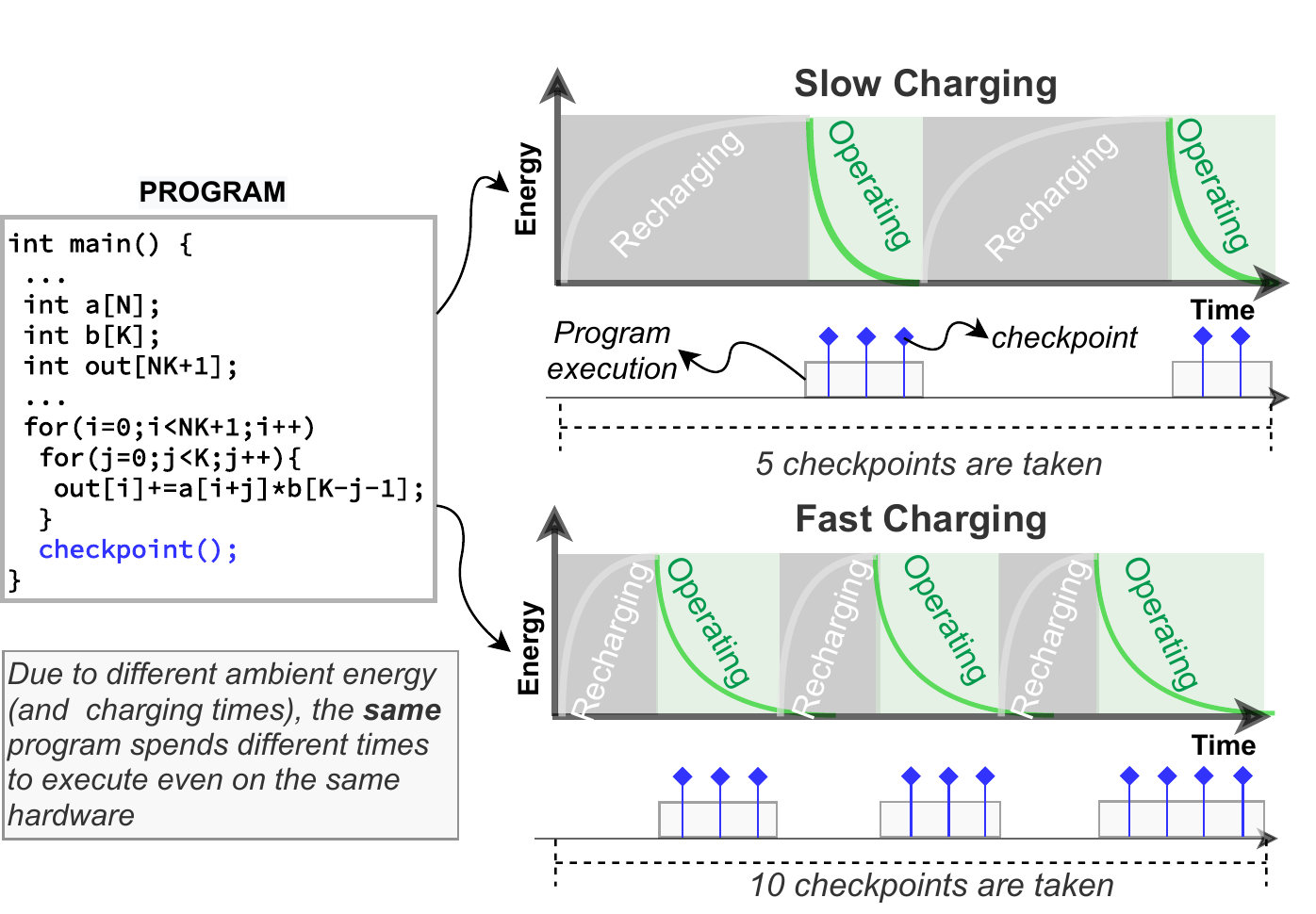}
    \caption{Dynamics of Intermittent Execution. A sample application composed of matrix multiplication in a nested loop. The run-time behavior of the intermittent program (e.g., execution time) depends on the environmental factors (e.g., ambient energy), checkpoint placement and overhead, and the target platform (e.g., power consumption and capacitor size).}
    \label{fig:background}
\end{figure}

\subsection{The Need for Timing Behavior Analysis}

Using existing intermittent programming models, programmers mainly focus on the progress, memory consistency, and functional correctness of their programs.  
Without analyzing the timing behavior affected by several factors, programmers cannot be sure if their intermittent programs meet throughput expectations in a deployment environment. 

Figure~\ref{fig:background} depicts how the execution time of a checkpointed code changes under different environmental conditions. In high-energy environments, the capacitor charges faster, the program progresses quickly, the checkpoints are triggered more frequently, and the execution needs a shorter time. The same program takes much longer in low-energy environments. The device is mostly off to charge its capacitor and becomes active only for a short time to perform computation. 

Some runtime environments, e.g.,~\cite{mayfly, ink, tics}, support the expression of timeliness in program source and check if sensing and computation are handled timely \emph{at runtime}. They can catch the sensor readings that lost their timeliness due to power failures and long charging times, throw away these values, or omit their computation. Even though catching data and computation expiration is beneficial, programmers have no opinion, \emph{at compile-time}, on how their programs behave. For instance, sensor readings and computations might expire continuously in specific energy harvesting conditions due to a wrong checkpoint placement or insufficient capacitor size. The runtime environments can catch these expirations to prevent unnecessary processing and save precious harvested energy. However, the program does not produce any meaningful results due to frequent data expirations. 
With timing behavior analysis, programmers will know, in advance, how their programs will execute on the target environment (e.g., if there will be frequent data expirations) and perform the necessary actions to let their programs meet their expectations.

\subsection{Factors Affecting Timing Behavior}



\subsubsection*{Energy Harvesting Environment.} The availability of ambient energy sources is unpredictable. The harvested energy depends on several factors, such as the energy source type (e.g., solar or radiofrequency), distance to the energy source, and the efficiency of the energy harvesting circuit. A probabilistic model of the energy harvesting environment can be derived based on observations and profiling~\cite{preact,mishra2015charging, naderi2015wireless}. When incoming power is strong enough, the capacitor charges rapidly, and the device becomes available quickly after a power failure. At low input power, the charging is slower and takes more time. 

\subsubsection*{Energy Storage.} The interaction between the capacitor and the processor in battery-free devices plays a crucial role in the program execution time. When the capacitor is fully charged, the device turns on and starts program execution. When the stored energy in the capacitor drops below a threshold, the device dies. One of the factors that affect the device on time is the capacitor size. If the capacitor size is large, the device has more energy to spend until the power failure, but charging the capacitor takes more time. Prior works, e.g.,~\cite{mishra2015charging}, proposed models that capture the charging behavior of capacitors.

\subsubsection*{Target Platform.} The power requirements of the target platform affect the end-to-end delay of the program execution. A program might take a long time to finish on platforms with high power requirements since the capacitor discharges faster. Hence, the program might drain the capacitor more frequently (since instructions consume more energy in a shorter time). Therefore, the device is interrupted by frequent power failures, and it is unavailable and charging its capacitor for long periods. 
We can model the target platform by using the instruction-level energy consumption profiles, as suggested in~\cite{epic, cleancut}.

\subsubsection*{Intermittent Runtime.} The programming model and the underlying runtime affect the execution time of intermittent programs. For instance, the checkpointing overhead is architecture-dependent since the number of registers and the volatile memory size change from target to target. Moreover, checkpoint placement is also crucial: the more frequent the checkpoints are, the more energy consumed, but less computation is lost upon a power failure. 

\subsubsection*{Program Inputs.} Intermittent program inputs are mostly the sensor readings that depend on the environmental phenomena. Different inputs lead to different execution flow, and in turn, energy consumption. The energy consumption affects the frequency of power failures and the charging time. 

\subsection{A Probabilistic Approach for Timing Analysis} 
The challenge is to devise a technique that facilitates the compile-time timing analysis of intermittent programs considering the factors mentioned above. Since these factors can be represented using stochastic models~\cite{mishra2015charging, naderi2015wireless, preact, epic, cleancut}, \emph{probabilistic symbolic execution}---a static analysis technique calculating the probability of executing parts of a program~\cite{geldenhuys2012probabilistic}---becomes an excellent fit for timing behavior analysis. 

Symbolic execution analyzes a program to discover which program inputs execute which program parts. It searches the execution tree of a program using symbolic values for program inputs. It generates from the execution tree program paths with a path condition, i.e., a conjunction of constraints on program inputs. When symbolic execution reaches a node of the execution tree, it evaluates the path condition describing the path from the root to that node. If the condition is satisfiable, it continues in that branch of the tree. If not, the branch is known to be unreachable. The output of satisfiability checking is either 0 or 1. The values 0 and 1 are rough approximations of the probability that the path condition is true. Probabilistic symbolic execution estimates finer approximations of the probability of path conditions being true~\cite{geldenhuys2012probabilistic}. One way is to count the number of solutions to a path condition (model counting~\cite{gomes2009model}) and divide it by the product of input domain. 


\section{\sysname: Systems Overview}
\label{sec:overview}

\begin{figure}
\begin{center}
\centerline{\includegraphics[trim={0.28cm 0.3cm 0.25cm 0.17cm},clip,width=0.75\columnwidth]{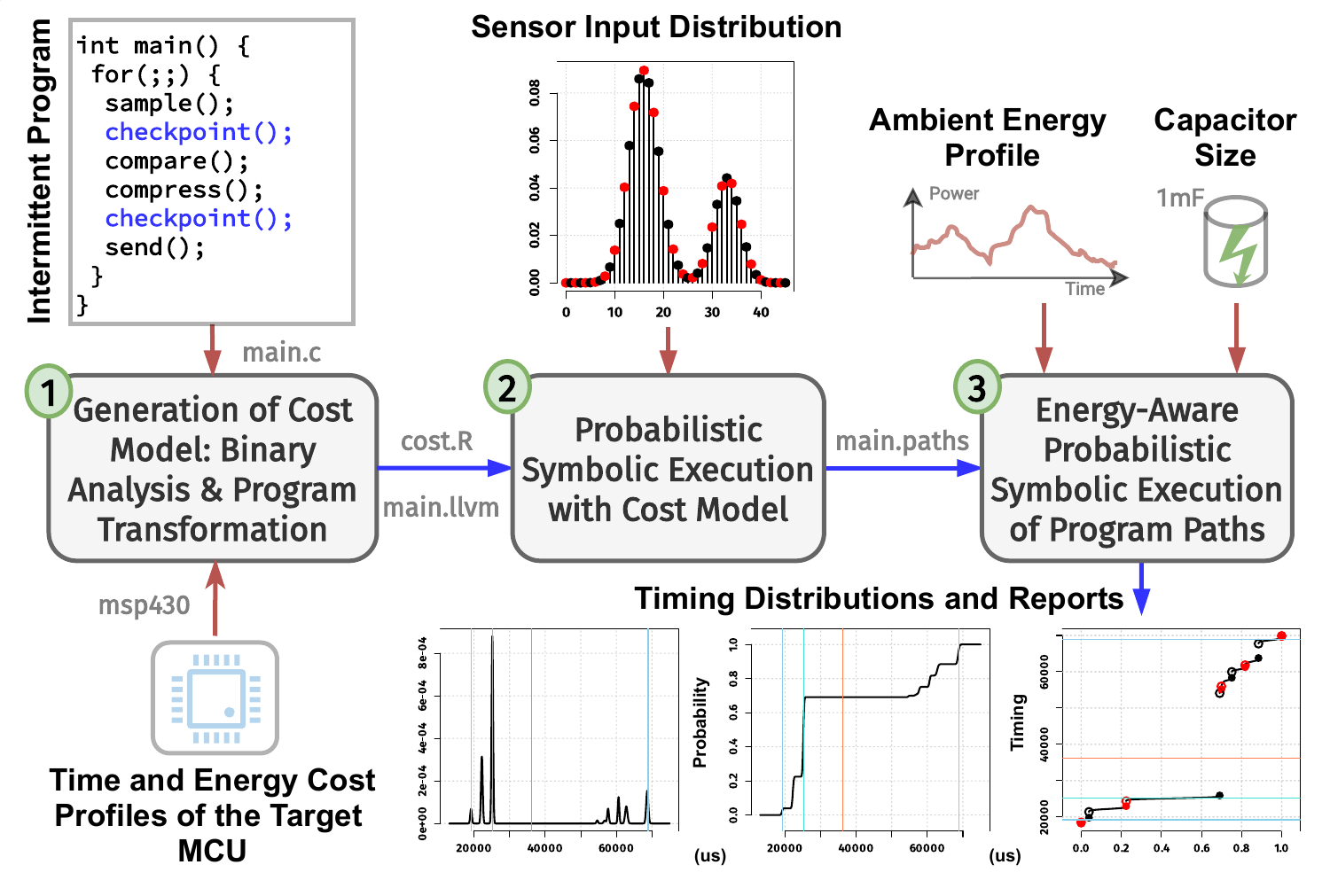}}
\caption{\sysname Overview.} \label{fig:approach}
\vspace*{-0.6cm} 
\end{center}
\end{figure}

The process in Figure~\ref{fig:approach} presents an overview of \sysname. Its steps are fully automated. Sections~\ref{sec:cost_models}-\ref{sec:energy_aware_analysis} provide details of each step. In Step 1, \sysname takes an intermittent program, and time and energy cost profiles of the target architecture (\textit{main.c} and \textit{msp430} in Figure~\ref{fig:approach}). It automatically generates a cost model with the program having split program blocks in LLVM IR~\cite{lattner2004llvm} (\textit{cost.R} and \textit{main.ll}). 
The cost model is an R specification having LLVM program blocks to which time and energy costs are assigned as probabilistic cost expressions. 

We designed our symbolic execution technique not to be specific to any microcontroller architecture and instruction set. Therefore, our symbolic execution runs not on microcontroller instructions but the program blocks in LLVM IR. 
This design choice contributes to the scalability of \sysname since generating paths on program block-level leads to fewer divergent (\emph{power failure paths}) to be analyzed. 
\sysname needs the time and energy cost profiles of the instruction types of each new target architecture platform organized under addressing modes. 
In Section~\ref{sec:evaluation}, we derived the cost profiles for the MSP430FR5994 platform~\cite{MSP430FR5994} 
through empirical data collection. It is a one-time effort for each new platform. Symbolic execution on program blocks may lead to coarse approximations of energy costs in the explored paths. Therefore, in Step 1, \sysname automatically splits blocks having outlier energy costs (maximum outliers) among all the program block energy costs. 

In Step 2, \sysname symbolically executes the program blocks in LLVM IR (\textit{main.ll}) with the cost model (\textit{cost.R}) to generate program paths and path execution probabilities (\textit{main.paths}) based on the sensor input distribution.
In Step 3, \sysname takes the ambient energy profile, capacitor size, program paths, and path execution probabilities as input. Each path is symbolically executed with stochastic ambient energy to estimate execution time probability distributions of program paths and functions for intermittent execution. 

\section{Specification of \sysname Inputs}
\label{sec:specification}

\begin{figure}
\begin{center}
\centerline{\includegraphics[width=.8\columnwidth]{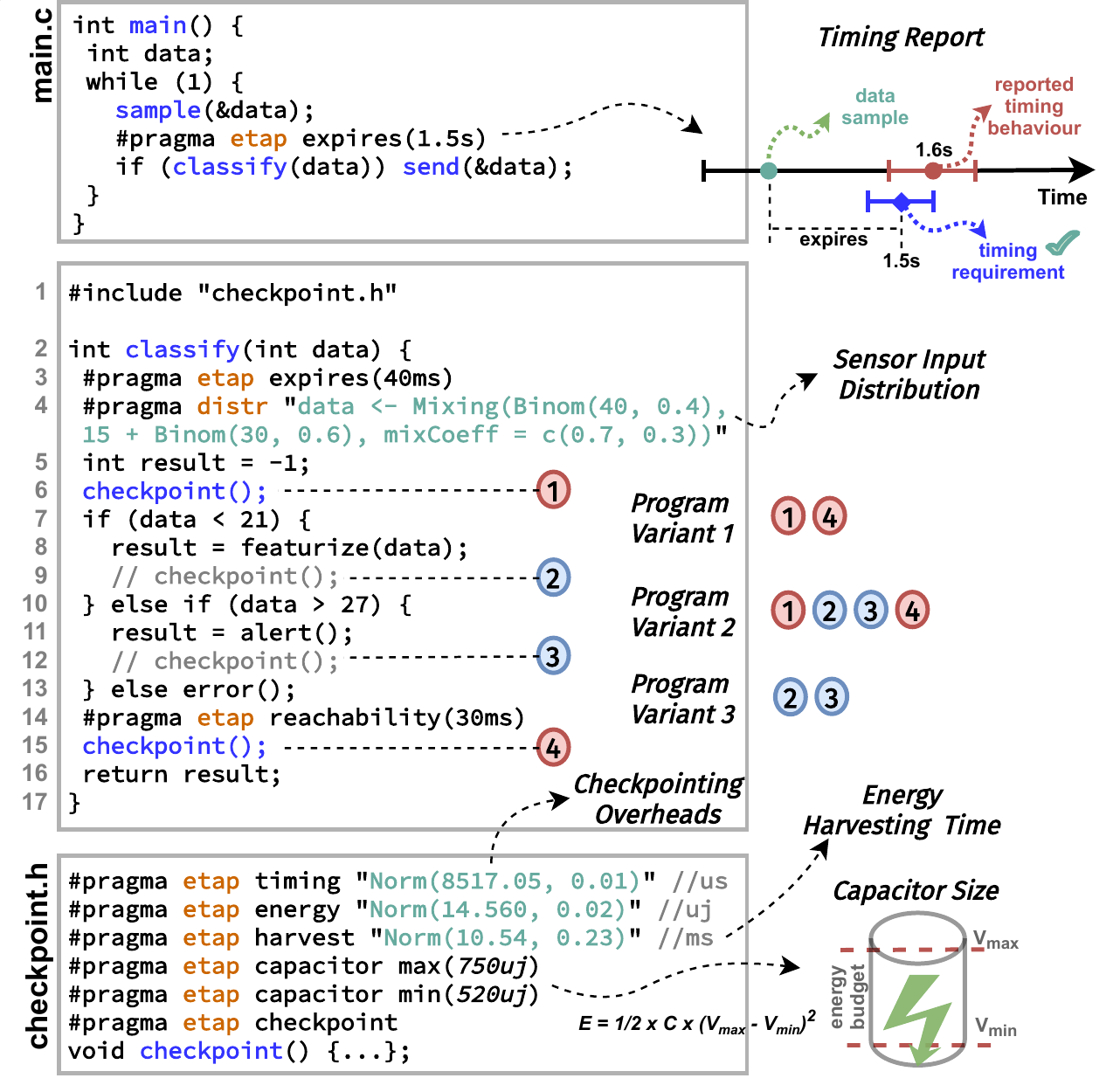}}
\caption{Example Intermittent Program with \sysname Pragmas.}
\label{fig:example}
\vspace*{-0.4cm} 
\end{center}
\end{figure}

Engineers specify \sysname inputs (e.g., timing requirements and probability distributions of sensor inputs) via the `\emph{\#pragma}' directive~\cite{Pragma}. Figure~\ref{fig:example} presents an example intermittent program (\textit{main.c}) and a header file (\textit{checkpoint.h}) for the checkpoint operation. The runtime environment provides the header file~\cite{tics}, which programmers extend with pragmas for the time and energy cost distributions of the checkpoint operation ($\tau_{cp}, E_{cp}$), ambient energy harvesting time profile ($\tau_{harvest}$) and capacitor size ($E_{min}, E_{max}$). 

The main function of the example program samples, classifies and sends the input data in a loop (\textit{main.c}); each loop iteration should execute in less than 1.5 second (
`\emph{\#pragma etap expires(1.5s)}' in \textit{main.c}). Function \textit{classify} interprets input data, alerts users, or reports an error (Lines 2-17). It should finish in less than 40 milliseconds (Line 3). It has two checkpoints (Lines 6 and 15); the second checkpoint should be reached from the first checkpoint in less than 30 milliseconds (`\emph{\#pragma etap reachability(30ms)}' in Line 14). The "reachability" pragma is to get the probability distribution of reaching at the specified checkpoint. The “expires” pragma is to get the probability distribution of timing at the specified point. The input distribution is discretized and modeled (`\emph{\#pragma distr}' in Line 4). It is the mixture of two signals following the binomial distribution (\emph{Binom(40, 0.4)} and \emph{Binom(30, 0.6)}). 

\section{Generation of Cost Models}
\label{sec:cost_models}
\sysname generates a cost model with the intermittent program that has split program blocks in LLVM IR (Step 1 in Figure~\ref{fig:approach}). 
It employs Clang~\cite{Clang} 
to compile the input program into the LLVM IR code. Each checkpoint call needs to be the first instruction in its basic block because the symbolic execution returns to the beginning of the basic block for a power failure. Therefore, \sysname splits checkpoint calls, which are not the first instruction, from their basic blocks.

The LLVM IR code is compiled into the assembly code for the target hardware architecture. \sysname performs a binary analysis to map the hardware instructions in the assembly code to the basic blocks. It calculates the time and energy cost distributions of each basic block. To do so, it convolves the cost distributions of all instructions in that block based on their types and addressing modes (see Table~\ref{tab:target-modeling}). Some basic blocks may need much more energy than other basic blocks, which may lead to coarse approximations of energy costs of the program paths. To have a better energy approximation, \sysname performs a semantic- and cost-preserving program transformation. It automatically splits blocks having outlier energy costs (i.e., maximum outliers) among all the program block energy costs for each function in the program. We use the IQR method~\cite{lock2020statistics} to detect the outlier blocks in terms of energy costs and obtain the threshold value. Considering the threshold, \sysname splits blocks having cost more than $Q_3 + 1.5(IQR)$ ($Q_3$ is the third quartile, and $IQR$ is the interquartile range). 

\begin{figure}[t!]
	\begin{center}
		\centerline{\includegraphics[width=.80\columnwidth]{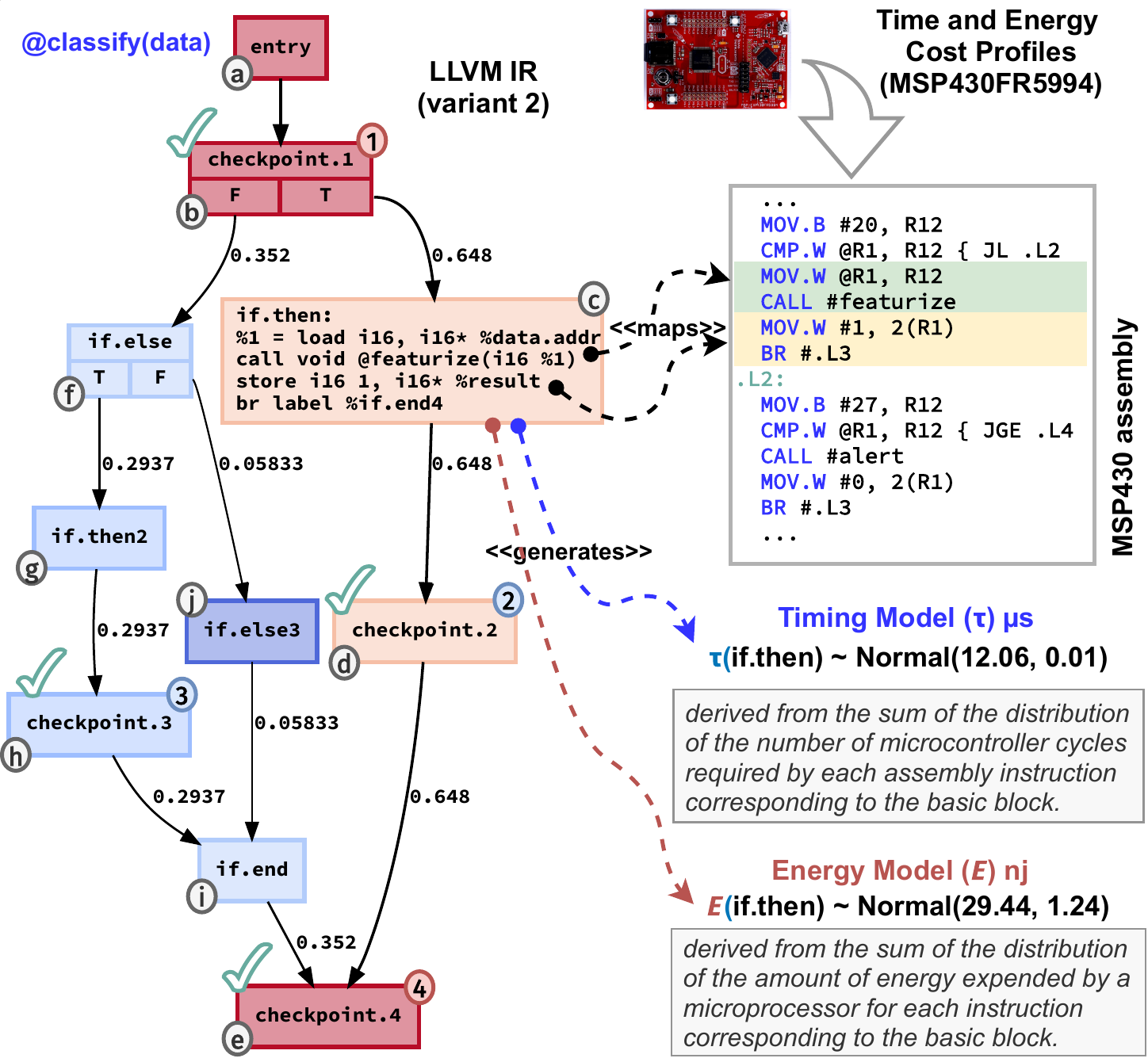}}
		\caption{CFG of Function \textit{classify} and its Cost Models.}
		\label{fig:cost-models}
  \vspace*{-0.4cm} 
	\end{center}
\end{figure}

Figure~\ref{fig:cost-models} presents the control flow graph (CFG) of \textit{classify} in Figure~\ref{fig:example} (with four checkpoints). Time and energy costs are modeled as normal distributions and encoded in an R specification (\textit{cost.R} in Figure~\ref{fig:approach}). Each node represents a basic block. Blocks \textit{a} and \textit{b}, \textit{c} and \textit{d}, and \textit{g} and \textit{h} were initially single blocks \sysname splits into two as the checkpoint calls were not the first instruction. 
No blocks are split for the threshold. 

\section{Probabilistic Path Exploration}
\label{sec:execution_cost_models}

\begin{figure*}
\begin{center}
\centerline{\includegraphics[width=1.07\linewidth]
{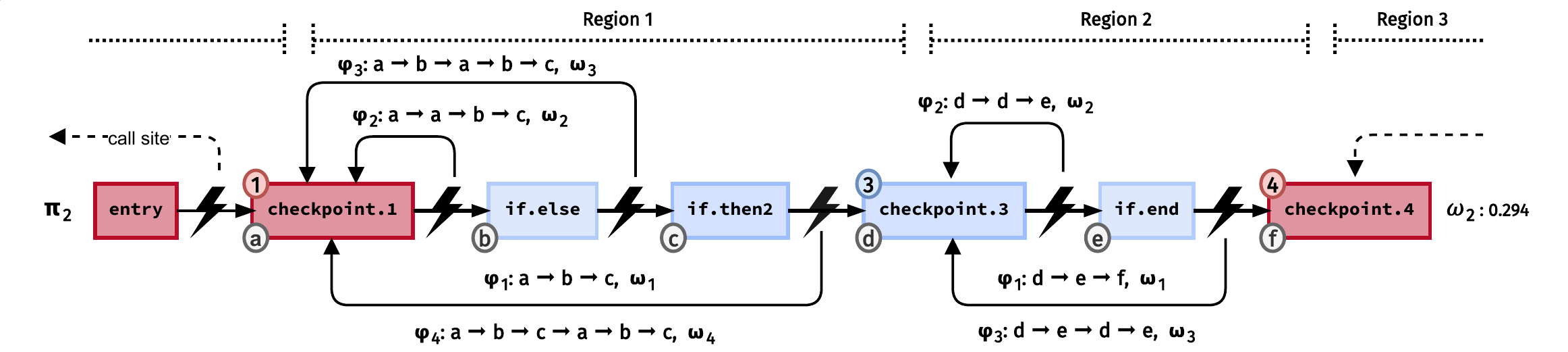}}
\caption{An Example Path derived for Normal Execution and its Analysis for Intermittent Execution.}
\label{fig:algorithm-instance}
\vspace*{-0.4cm} 
\end{center}
\end{figure*}

\sysname symbolically executes the basic blocks with the cost models in the R environment with depth-first search (Step 2 in Figure~\ref{fig:approach}). It generates program paths and their execution probabilities for \textit{program execution without power failures}. The program paths are later re-executed symbolically with stochastic ambient energy to estimate its execution time probability distribution for intermittent execution (see Section \ref{sec:energy_aware_analysis}). 
{\centering
\begin{minipage}{.6\linewidth}
\begin{algorithm}[H]
\DontPrintSemicolon
\small 

{\sc Blocks($path$)} $\gets$ {\sc Blocks($path$)} $\oplus$ {\sc Name}($curr$) \; 
{\sc Timing($path$)} $\gets$ {\sc Timing($path$)} $\ast$ {\sc Get}($Costs$, $curr$) \; 
{\sc Eval}({\sc Instructions}($curr$), $env$)\; 

\ForEach{$succ$ in {\sc Successors($curr$)}} { 
  $prob \gets prob$ $\times$ {\sc BranchProbability}($curr$, $succ$, $env$)\;
  \If{$prob > 0 \wedge \neg${\sc IsMaxLoopReached}} {
    $Paths \gets$ {\sc \textbf{Dfs}}($Paths$, $Costs$, $( b$, $succ$, $prob )$, {\sc CloneEnv($env$)}, $path$)
  }
}

\Return{$Paths$ $\oplus$ $($ {\sc Blocks($path$)}, $prob$, {\sc Timing($path$)} $)$} \; 

\caption{{\sc \textbf{DFS}} for Probabilistic Path Exploration}
\label{algo:dfs}
\end{algorithm}
\end{minipage}
\par
}



\noindent\sysname calculates the path execution probability and the execution time probability distribution for each program path in the function to be analyzed. 
Algorithm~\ref{algo:dfs} gives the depth-first search for probabilistic path exploration. 


\sysname adds the visited (current) basic block of the CFG to the path (Line 1). It convolves the execution time probability distribution of the path and the execution time cost distribution of the visited block (Line 2). The block instructions in the path environment are evaluated to decide which successor block to take in which branching conditions (Line 3). 
\sysname models path execution environments with instances of a symbolic memory model in which LLVM instructions are probabilistically interpreted. For instance, in the LLVM instruction `\emph{$\%inc$ = add nsw i16 $\%$i.0, 1}', if `\emph{$\%$i}' is a random variable (r.v.) that follows a discrete probability distribution, the 16-bit integer register $\%inc$ becomes an r.v., \emph{$\%inc \sim \;\%$i.0 $+ 1$}, through linear location-scale transformation~\cite{blitzstein2019introduction}. If the instruction was among two registers, `\emph{$\%inc$ = add nsw i16 $\%$i.0, $\%$i.1}' and in that `\emph{$\%$i.0}' and `\emph{$\%$i.1}' are independent r.v.s, the expression would be interpreted as the sum (integral) of two r.v.s (convolution)~\cite{blitzstein2019introduction}.
The path execution probability is a conditional probability and recalculated for each new successor block. The new path probability is the multiplication of the current path probability and the branch probability for the successor basic block (Line 5). Figure~\ref{fig:cost-models} shows the probabilities in the example CFG. The paths are recursively explored until the path probability is equal to zero or the maximum number of iterations for a loop is reached (Line 6-7). 

LLVM operators on distributions are overloaded to produce transformed distributions. When symbolic execution has reached a branch point, \sysname integrates the probability function of the transformed distribution that the execution branches on over the region of interest. This tells us the probability of taking that branch. After stepping over a branch point, we perform state forking, similar to mainstream symbolic execution engines~\cite{klee,symexsurvey}, and update the random variables in the symbolic environment of the path with the information gained. 

Linear transformations of named distributions have mostly analytical solutions; however, we need to employ computational approaches if the transformation is non-linear (e.g., a logarithmic transformation). This is also the case for convolution operations. For instance, the sum of two independent normally distributed (Gaussian) random variables also follows a normal distribution, with its mean being the sum of the two means, and its variance being the sum of the two variances; on the other hand, the sum of two random variables, which follow Uniform and Normal distributions respectively, cannot be expressed with a closed-form expression, and therefore, in these cases, \sysname approximates these probabilities using numerical methods.



For function \textit{classify} in Figure~\ref{fig:cost-models}, \sysname explores three paths i.e., $\pi_1$: `$a \rightarrow b(cp) \rightarrow c \rightarrow d(cp) \rightarrow e(cp)$', $\pi_2$: `$a \rightarrow b(cp) \rightarrow f \rightarrow g \rightarrow h(cp) \rightarrow i \rightarrow e(cp)$', and $\pi_3$: `$a \rightarrow b(cp) \rightarrow f \rightarrow j \rightarrow i \rightarrow e(cp)$' where $cp$ indicates the blocks with the checkpoint operation. It calculates path execution probabilities (or weights) associated with each path ($\omega_1 = 0.648$, $\omega_2 = 0.058$, and $\omega_3 = 0.294$). The nodes labeled with small letters represent the basic blocks (e.g., $a$, $b$, $c$). Each basic block is heat-map colored based on its execution probability. 
Each path has an execution time probability distribution ($\tau_1$, $\tau_2$, and $\tau_3$) that represents the path's timing behavior while it is continuously powered. 
It is the convolution of the execution time cost distribution of the basic blocks along the path. Since the timing behavior is \textit{compositional} when there is no power failure, \sysname considers checkpoint ($\tau_{cp}$) and other function calls in the program path ($\tau_{alert}$, $\tau_{featurize}$, and $\tau_{error}$) while convolving the cost distributions as follows:
{
\begin{align*}
&\tau_1 \sim a(\tau) \ast b(\tau) \ast \tau_{cp} \ast c(\tau) \ast \tau_{featurize} \ast d(\tau) \ast \tau_{cp} \ast e(\tau) \ast \tau_{cp} 
\\
&\tau_2 \sim a(\tau) \ast b(\tau) \ast \tau_{cp} \ast f(\tau) \ast g(\tau) \ast \tau_{alert} \ast h(\tau) \ast \tau_{cp} \ast i(\tau) \ast e(\tau) \ast \tau_{cp} 
\\
&\tau_3 \sim a(\tau) \ast b(\tau) \ast \tau_{cp} \ast f(\tau) \ast j(\tau) \ast \tau_{error} \ast i(\tau) \ast e(\tau) \ast \tau_{cp}
\end{align*}
}
The execution time probability distribution of function \textit{classify} for program execution without power failures is a convex combination of all the timing distributions of its execution paths: $\tau_{classify} \sim \omega_1\tau_1 + \omega_2\tau_2 + \omega_3\tau_3$. It is a univariate mixture distribution and represents the timing behavior of the function while the energy is plentiful in the environment. 

\section{Energy-aware Analysis of Program Paths}
\label{sec:energy_aware_analysis}

\begin{figure*}[t!]
\centering
\hspace{-0.03cm}\llap{\raisebox{0cm}{\includegraphics[height=0.8cm]{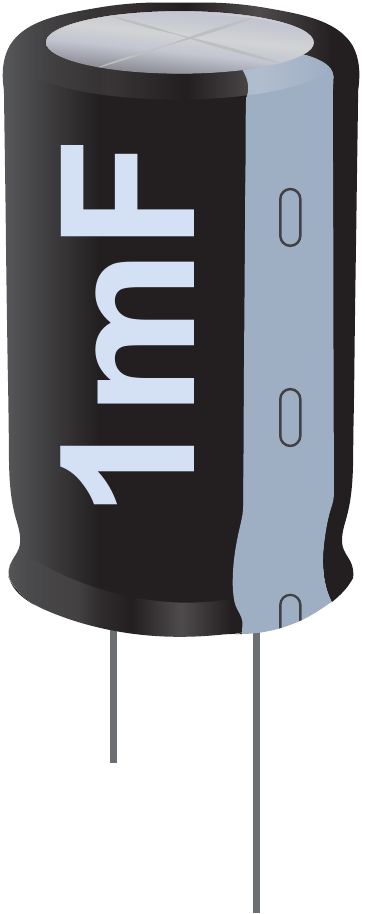}}}
\subfloat[1mF Capacitor - Program Variant 1]
{\label{fig:a}\includegraphics[width=0.32\linewidth, trim={0.6cm 1.3cm 0.6cm 1cm}, clip]{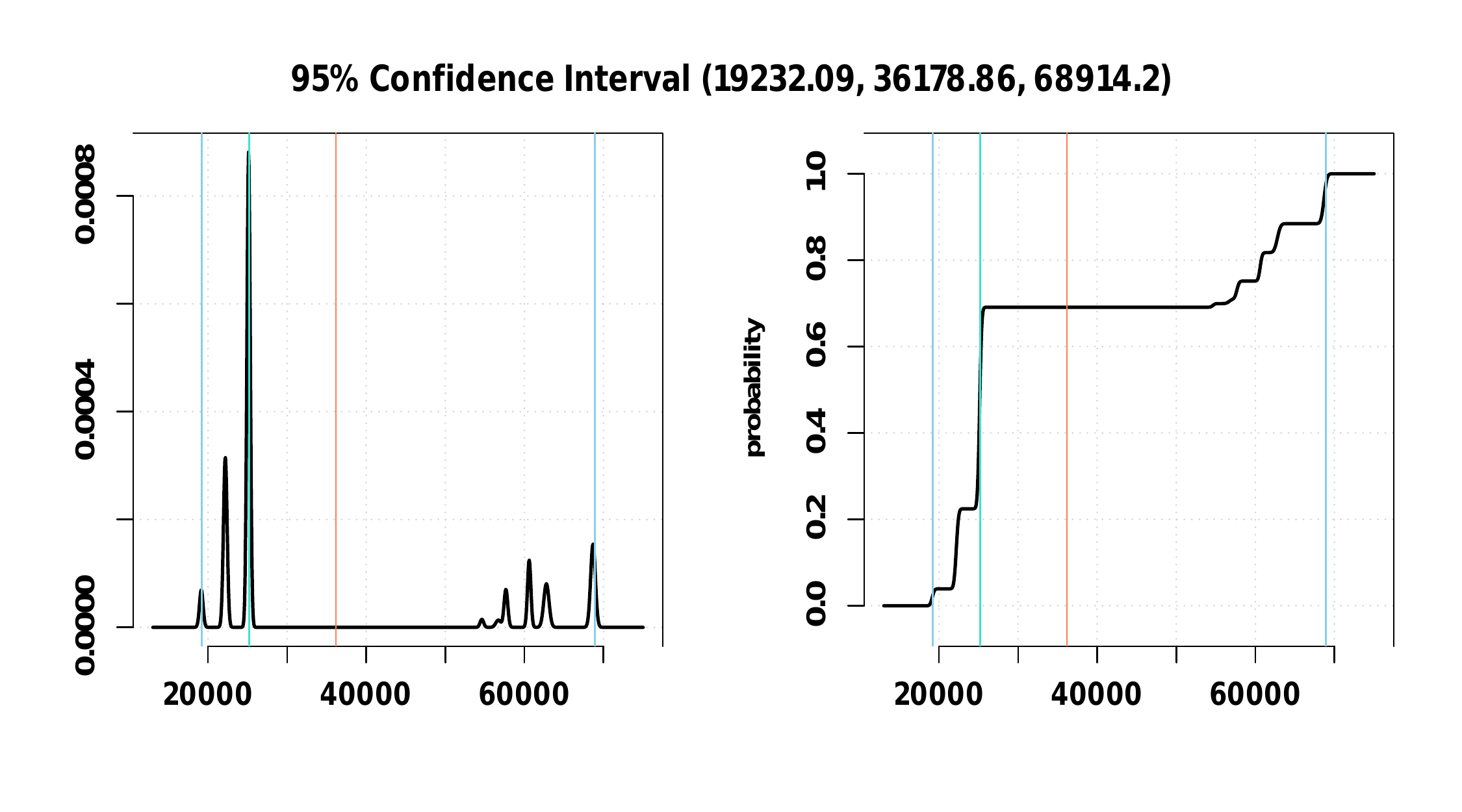}}
\hspace{-0.09cm}\llap{\raisebox{2.4cm}{\includegraphics[height=.48cm]{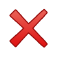}}}
\quad
\subfloat[1mF Capacitor - Program Variant 2]
{\label{fig:b}\includegraphics[width=0.32\linewidth, trim={0.6cm 1.3cm 0.6cm 1cm}, clip]{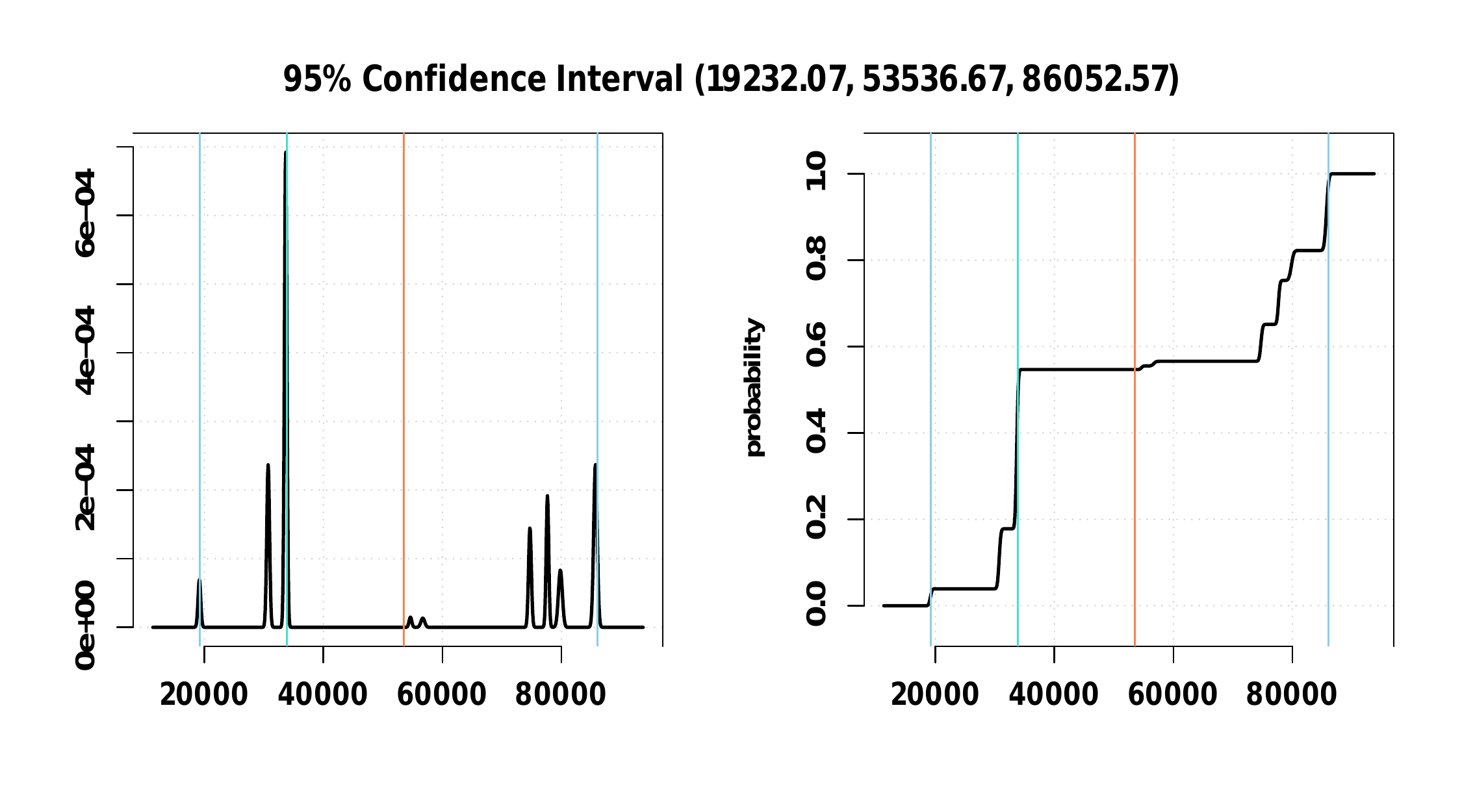}}
\hspace{-0.09cm}\llap{\raisebox{2.4cm}{\includegraphics[height=.48cm]{images/nope-asplos.pdf}}}
\quad
\subfloat[1mF Capacitor - Program Variant 3] 
{\label{fig:c}\includegraphics[width=0.32\linewidth, trim={0.6cm 1.3cm 0.6cm 1cm}, clip]{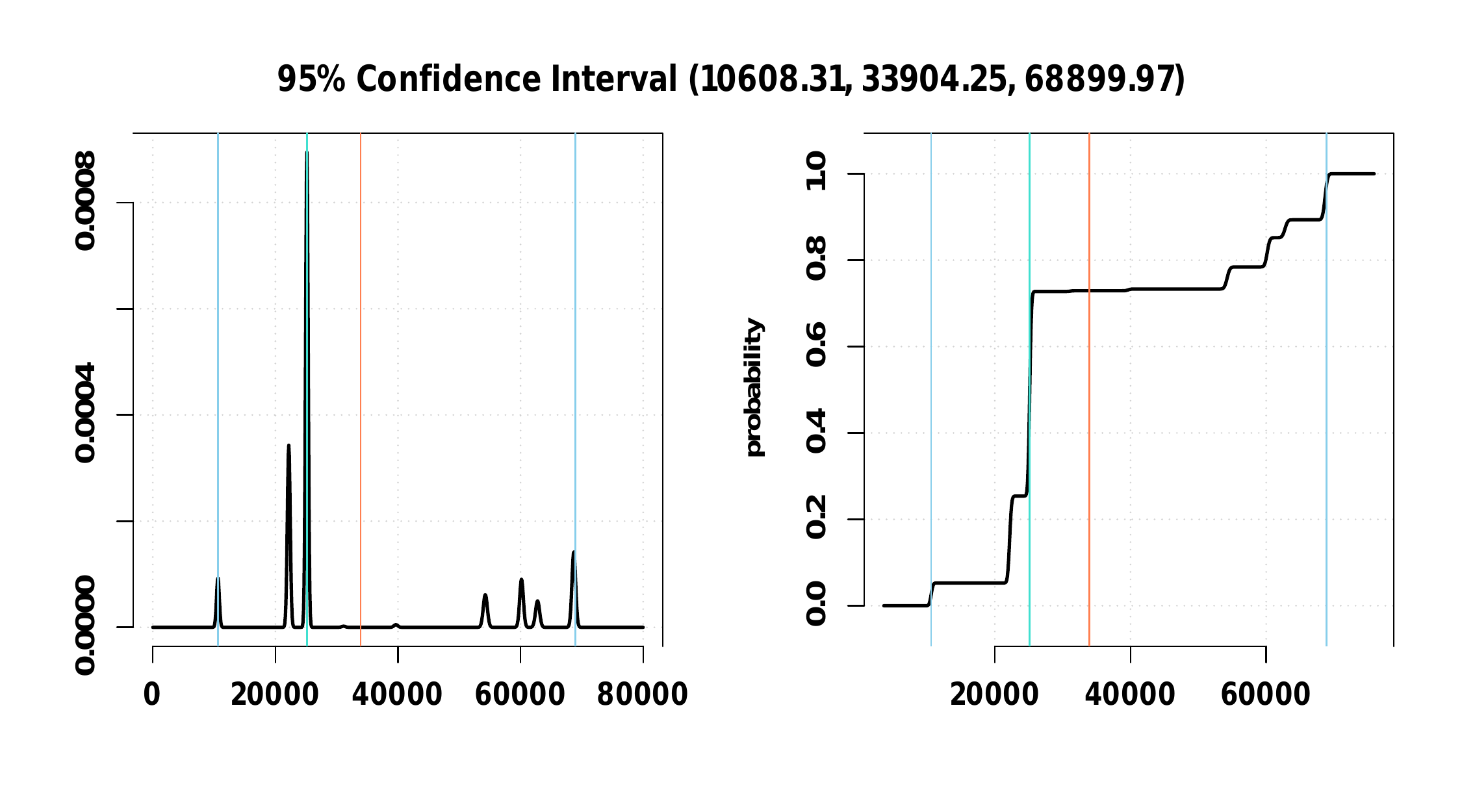}}
\hspace{-0.09cm}\llap{\raisebox{2.4cm}{\includegraphics[height=.48cm]{images/nope-asplos.pdf}}}
\\
\hspace{-0.03cm}\llap{\raisebox{0cm}{\includegraphics[height=0.9cm]{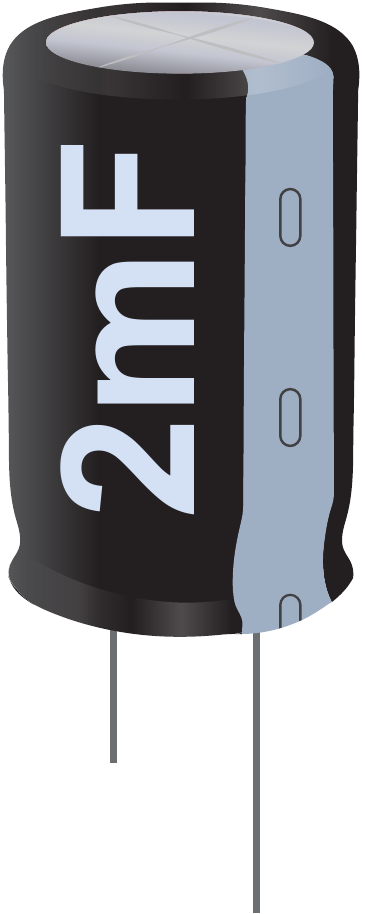}}}
\subfloat[2mF Capacitor - Program Variant 1]
{\label{fig:d}\includegraphics[width=0.32\linewidth, trim={0.6cm 1.3cm 0.6cm 0.9cm}, clip]{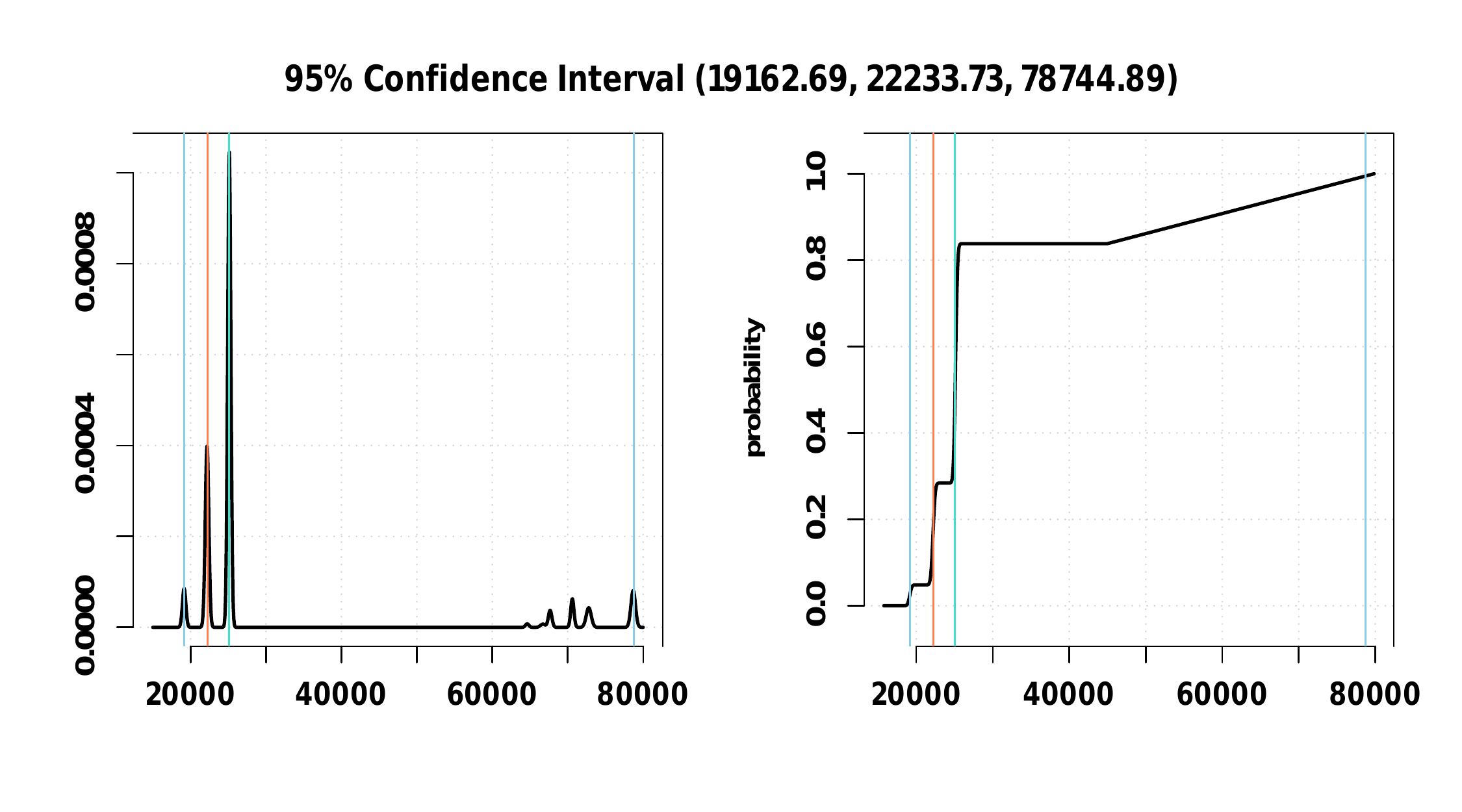}}
\hspace{-0.08cm}\llap{\raisebox{2.5cm}{\includegraphics[height=.45cm]{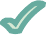}}}
\quad
\subfloat[2mF Capacitor - Program Variant 2]
{\label{fig:e}\includegraphics[width=0.32\linewidth, trim={0.6cm 1.3cm 0.6cm 0.9cm}, clip]{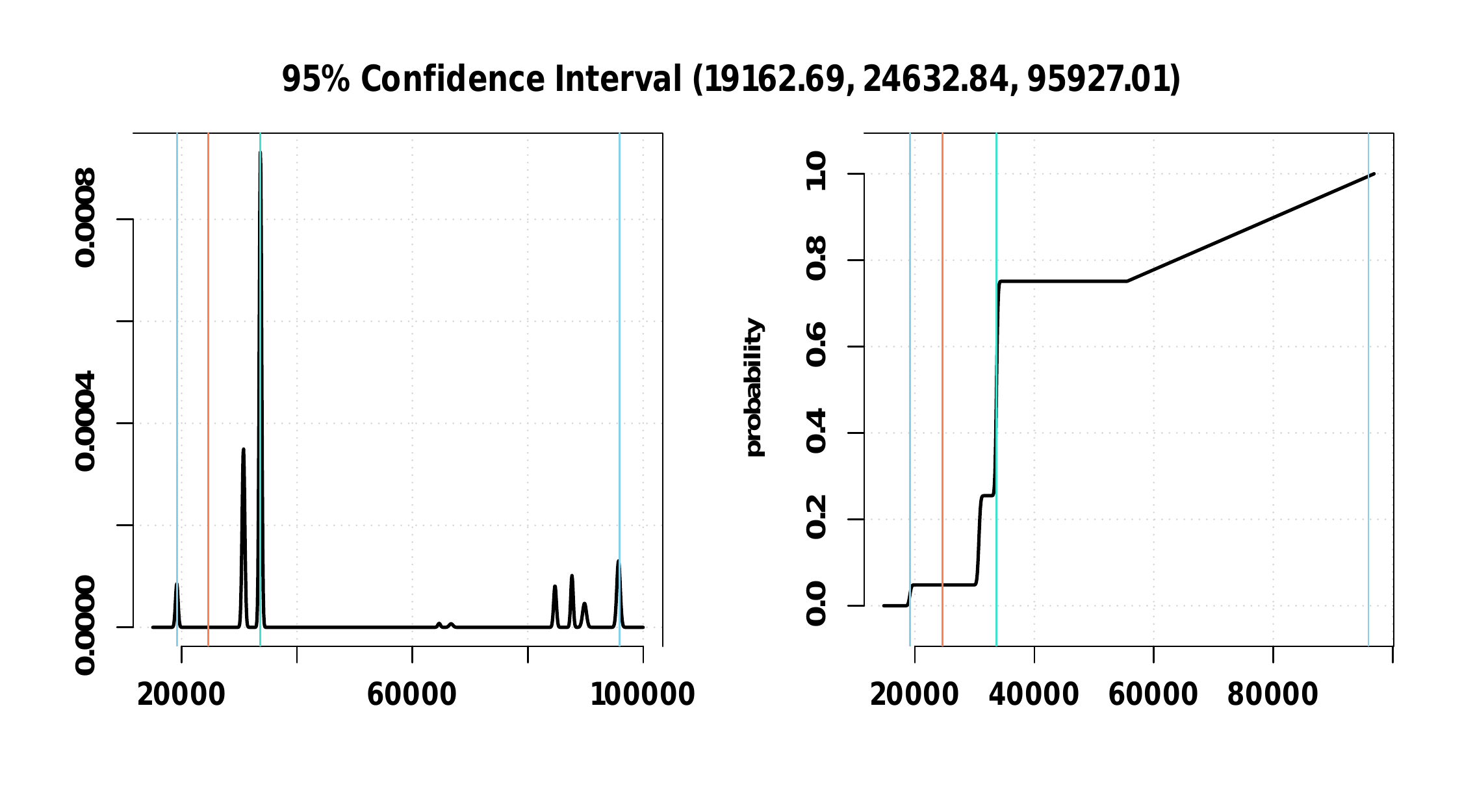}}
\hspace{-0.09cm}\llap{\raisebox{2.4cm}{\includegraphics[height=.48cm]{images/nope-asplos.pdf}}}
\quad
\subfloat[2mF Capacitor - Program Variant 3]
{\label{fig:f}\includegraphics[width=0.32\linewidth, trim={0.6cm 1.3cm 0.6cm 0.9cm}, clip]{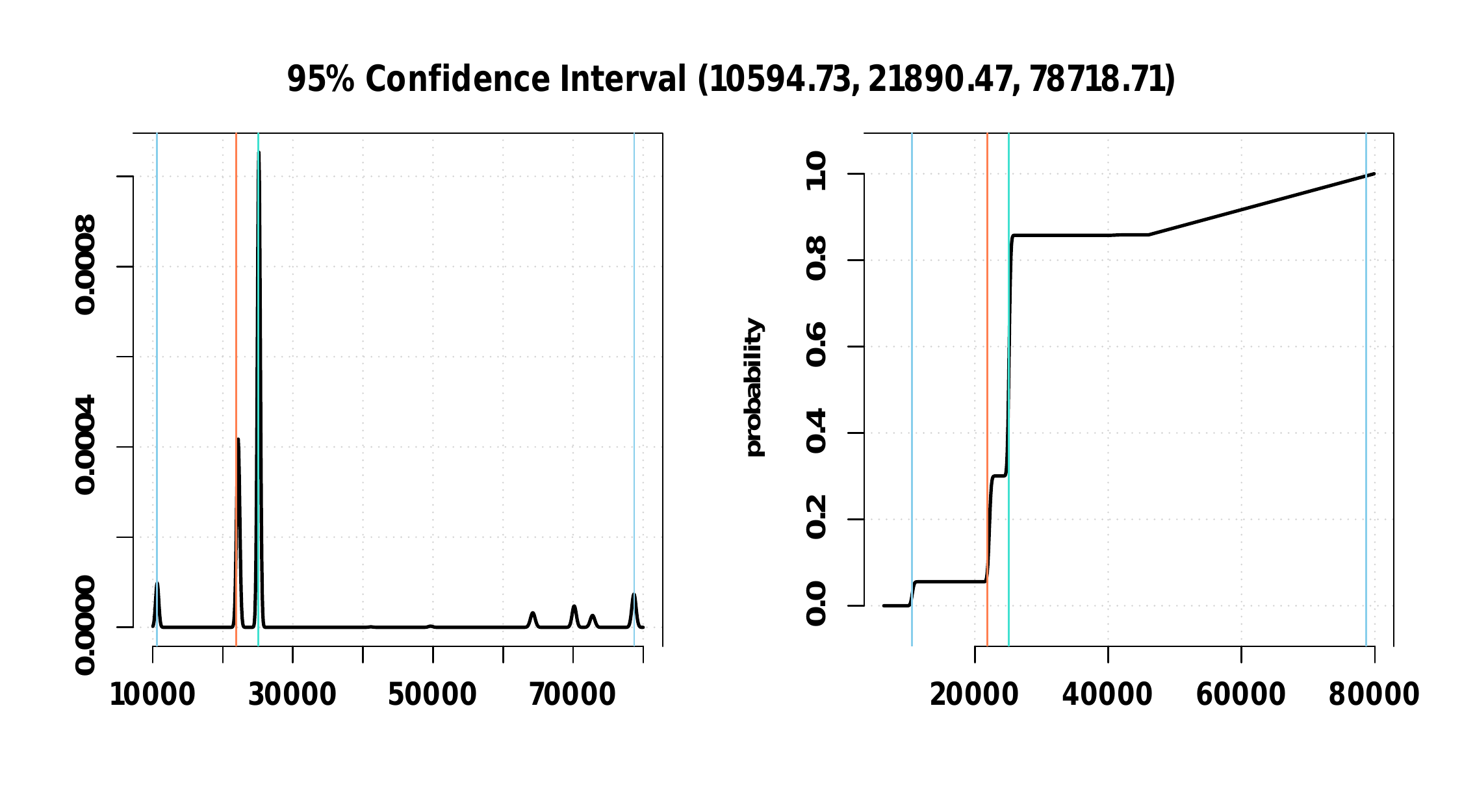}}
\hspace{-0.08cm}\llap{\raisebox{2.5cm}{\includegraphics[height=.5cm]{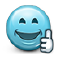}}}
\\
\hspace{-0.035cm}\llap{\raisebox{0cm}{\includegraphics[height=1cm]{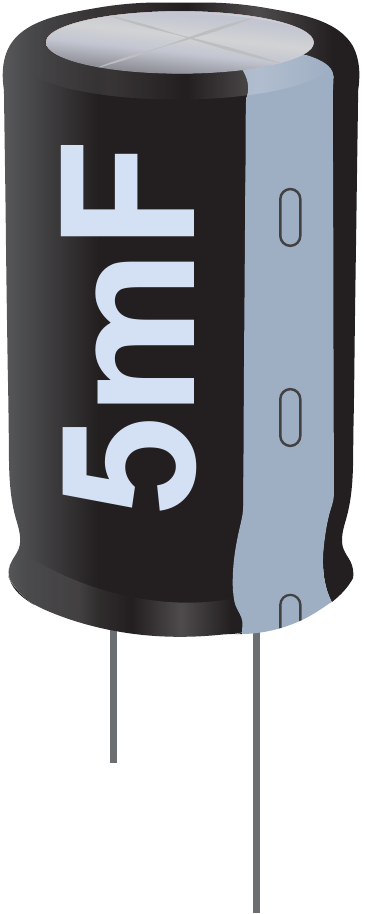}}}
\subfloat[5mF Capacitor - Program Variant 1]
{\label{fig:g}\includegraphics[width=0.32\linewidth, trim={0.6cm 1.3cm 0.6cm 0.9cm}, clip]{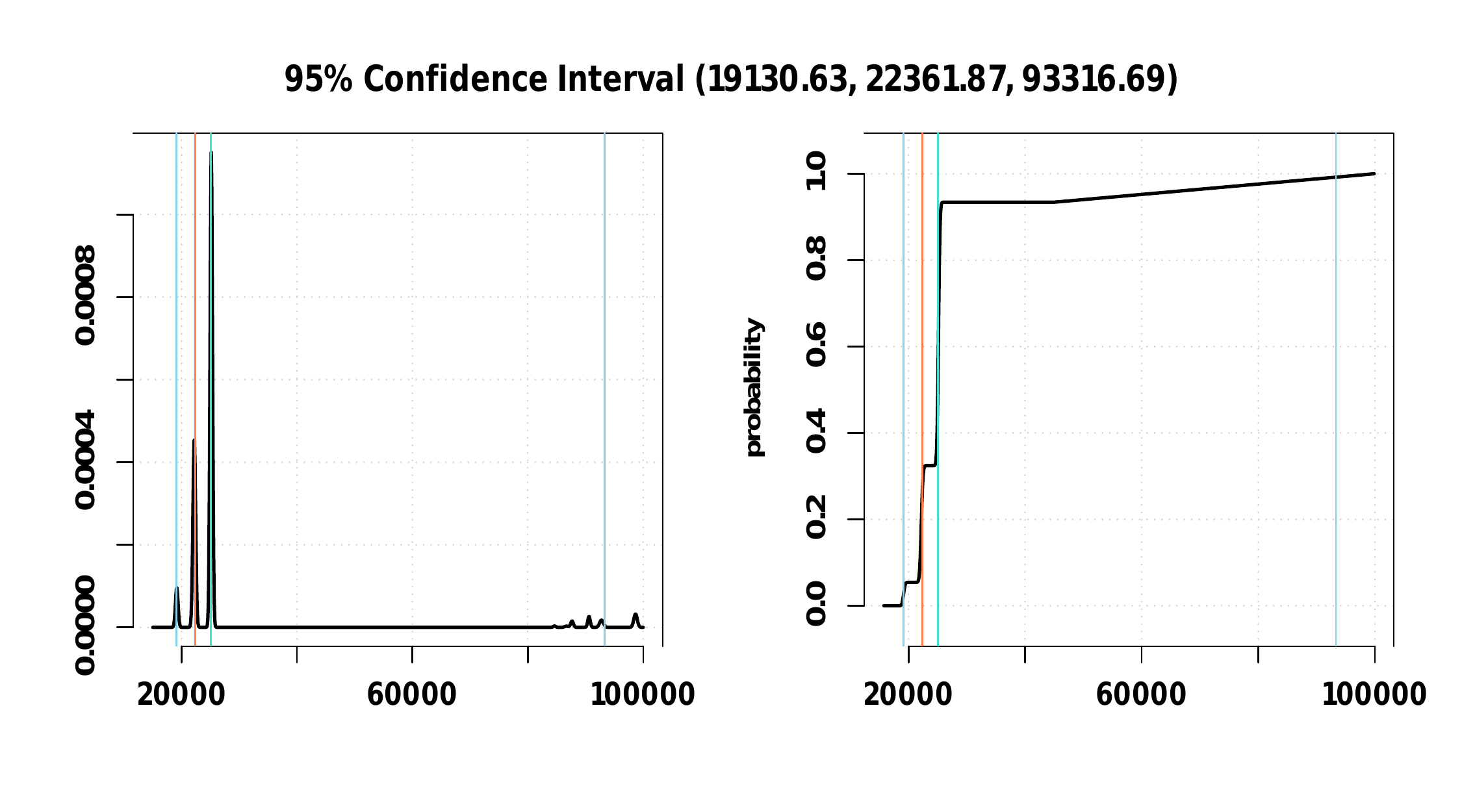}}
\hspace{-0.08cm}\llap{\raisebox{2.5cm}{\includegraphics[height=.4cm]{images/example_variant_selected.pdf}}}
\quad
\subfloat[5mF Capacitor - Program Variant 2]
{\label{fig:h}\includegraphics[width=0.32\linewidth, trim={0.6cm 1.3cm 0.6cm 0.9cm}, clip]{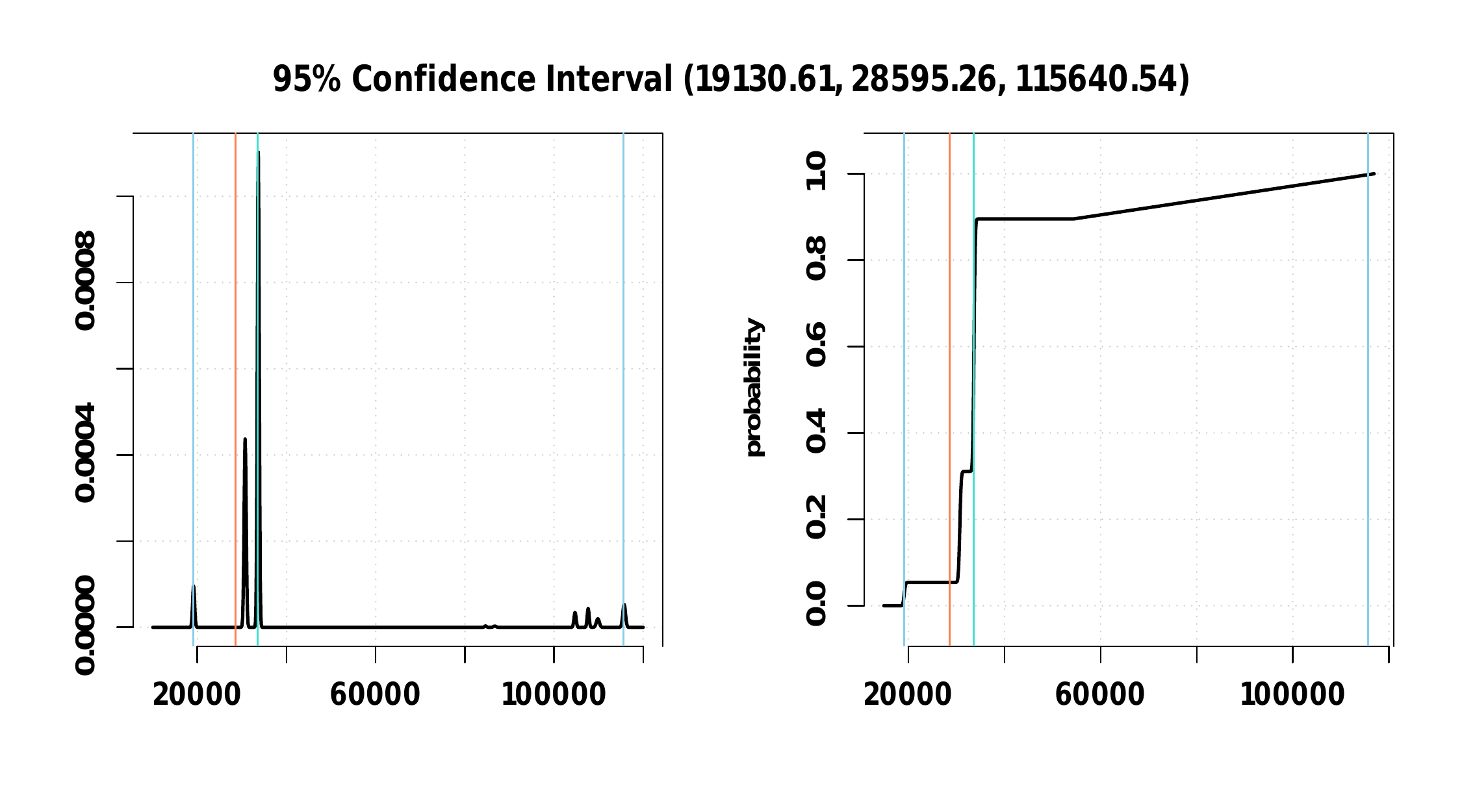}}
\hspace{-0.08cm}\llap{\raisebox{2.5cm}{\includegraphics[height=.4cm]{images/example_variant_selected.pdf}}}
\quad
\subfloat[5mF Capacitor - Program Variant 3]
{\label{fig:i}\includegraphics[width=0.32\linewidth, trim={0.6cm 1.3cm 0.6cm 0.9cm}, clip]{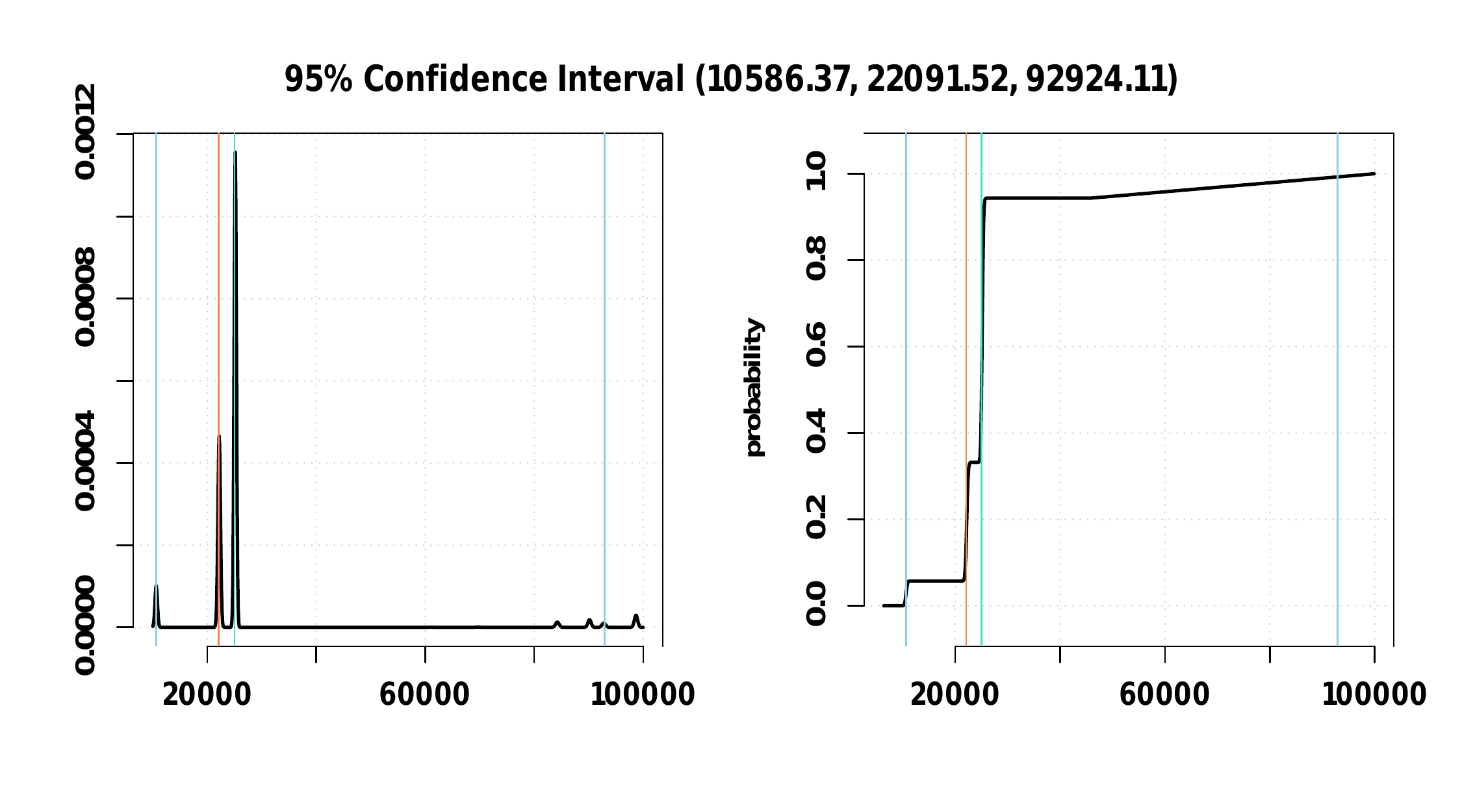}}
\hspace{-0.08cm}\llap{\raisebox{2.5cm}{\includegraphics[height=.4cm]{images/example_variant_selected.pdf}}}

\caption{Reports of the Execution Time Probability Distributions (PDF and CDF) of Function \textit{classify} after symbolically executing with 1MHz clock-rate under 3 different capacitor type and 3 different checkpoint placements with the given $\mathbf{\tau_{harvest}}$ profiles.}
\label{fig:output-intermittent-execution}
\end{figure*}

\sysname symbolically executes each program path with stochastic ambient energy to estimate its execution time probability distribution for intermittent execution (Step 3 in Figure~\ref{fig:approach}). 
Figure~\ref{fig:algorithm-instance} presents an example path without power failures ($\pi_2$: `$a(cp) \rightarrow b \rightarrow c \rightarrow d(cp) \rightarrow e \rightarrow f(cp)$' where $cp$ indicates the blocks with the checkpoint operation) and illustrates its energy-aware symbolic analysis. We relabel the blocks in path $\pi_2$ to increase the readability in Figure~\ref{fig:algorithm-instance}. We also initiate the symbolic analysis from block $a$ to simplify the presentation. 

The analysis starts with a non-deterministic initial capacitor energy budget, i.e., a uniform random variable on the interval of the capacitor's maximum and minimum thresholds, $E_{init} \sim \uniform(E_{min}, E_{max})$. \sysname divides the path into regions between two sequential checkpoints and between the beginning/end of the path and the first/last checkpoint in the path (`$a \rightarrow b \rightarrow c$', `$d \rightarrow e$', and `$f$'). 
The path is analyzed region by region. The power failure probabilities are calculated for each block in the first region based on the initial capacitor energy distribution ($E_{init}$). If a power failure is probable, \sysname creates a path which returns to the checkpoint block. In addition to the initial program path in the first region in Figure~\ref{fig:algorithm-instance} ($\varphi_1$: `$a \rightarrow b \rightarrow c$'), three new paths are derived for power failures (e.g., $\varphi_3$: `$a \rightarrow \dot{b} \rightarrow a \rightarrow b \rightarrow c$' where the power failure occurs in block $b$). Each new path has only one power failure because \sysname reports \emph{non-terminating} ones~\cite{cleancut} that need more energy than the capacitor size. 

For each path, \sysname subtracts the energy cost distributions of program blocks from the capacitor energy through convolution; it sums (convolves) the time cost distributions of the blocks. The result is the time and energy cost distributions of the paths in the first region ($\tau_{1}$, $\tau_{2}$, $\tau_{3}$, $\tau_{4}$, $E_{1}$, $E_{2}$, $E_{3}$, and $E_{4}$). To reduce the number of convolution operations in the next region, \sysname derives univariate mixture distributions (e.g., $\tau_{mix_1}$) from the cost distributions of the paths with power failure (e.g. in region 1, $\tau_{mix_1} \sim (\omega_2 \tau_{\varphi_{2}} \ast \omega_3 \tau_{\varphi_{3}} \ast \omega_4 \tau_{\varphi_{4}}) \ast \tau_{harvest}$, where $\omega_2$, $\omega_3$, and $\omega_4$ are failure probabilities). These distributions are input time costs for the paths in the next region derived from the power failure paths in the current region (e.g., $\varphi_{2} \sim \tau_{mix_1} \ast d_{\tau} \ast (d_{\tau} \ast e_{\tau})$ for $\varphi_2$ in the second region in Figure~\ref{fig:algorithm-instance}). 

The current capacitor energy distribution is calculated for each path in the first region. For the capacitor energy of the path without power failure, \sysname subtracts the energy cost distributions of the blocks from the initial capacitor energy distribution through convolution, i.e., $E_{current} \sim E_{init} \ast - (a_{\epsilon} \ast b_{\epsilon} \ast c_{\epsilon})$. In the power failure paths, the capacitor is charged up to the maximum threshold energy level during power failure (sleep mode). Therefore, while calculating their capacitor energy, \sysname considers the maximum energy the capacitor can store, e.g., $E_{current} \sim E_{max} \ast - (a_{\tau} \ast b_{\tau}  \ast a_{\tau} \ast b_{\tau} \ast c_{\tau})$ for path $\varphi_3$: `$a \rightarrow b \rightarrow a \rightarrow b \rightarrow c$'.
It uses the final program paths, capacitor energy levels, and cost distributions of the \textit{current region} as the initial parameters of the \textit{next region}. The same mixture and convolution operations are repeated in each new region. In the last region, the final energy and time probability distributions are obtained for the path under analysis (see \textit{Region 3} in Figure~\ref{fig:algorithm-instance}). 
 
As we described above, \sysname obtains the execution time probability distributions of each path of the function under analysis. These distributions are also used to generate reports on how likely timing requirements are met (see the 2\textsuperscript{nd} column in Table~\ref{tab:timing-report}). 
\sysname symbolically executes each path of the function in Figure~\ref{fig:cost-models} for intermittent execution. The mixture distribution of the execution time distributions of these paths is the execution time probability distribution of the function. 

Figure~\ref{fig:output-intermittent-execution} shows the reports of the execution time probability distributions of function \textit{classify} for intermittent execution under three different capacitor types (1mF, 2mF, and 5mF) and three different checkpoint placements (variant 1, variant 2, and variant 3 in Figure~\ref{fig:example}). The first functions in Figure~\ref{fig:output-intermittent-execution} are the probability density function (PDF), and the second ones are the cumulative distribution function (CDF). As shown in Table~\ref{tab:timing-report}, \sysname also generates a report for each configuration. 
\begin{table}
	\centering
	\caption{Timing Report of `\emph{\#pragma etap expires(40ms)}' for the given configuration space in Figure~\ref{fig:output-intermittent-execution}.}
	\label{tab:timing-report}
  \begin{tabularx}{0.6\linewidth}{@{}cccXXX@{}}
    \toprule
    Conf. &\multicolumn{1}{c}{Timing Probability}              &Meets  &\multicolumn{3}{c}{Timing Confidence Intervals (\textit{ms})} \\
    ~     &\multicolumn{1}{c}{$P_r(\tau_{classify} \le 40ms)$} &Req.?  &\multicolumn{1}{c}{\emph{95\% CI}} &\multicolumn{1}{c}{\emph{90\% CI}} &\multicolumn{1}{c}{\emph{80\% CI}} \\
    \midrule
    \rowcolor{black!3} 
    \subref{fig:a} &0.691 &\xmark &[19.2, 68.9] &[21.8, 68.7] &[22.1, 68.3]\\ 
    \subref{fig:b} &0.546 &\xmark &[19.2, 86.1] &[39.4, 85.9] &[30.7, 85.7]\\ 
    \rowcolor{black!3} 
    \subref{fig:c} &0.732 &\xmark &[10.6, 68.9] &[11.0, 68.7] &[22.0, 68.2]\\ 
    \midrule
    \subref{fig:d} &0.838 &\cmark &[19.2, 78.7] &[21.6, 78.4] &[22.0, 70.9]\\ 
    \rowcolor{black!3} 
    \subref{fig:e} &0.751 &\xmark &[19.2, 95.9] &[30.2, 95.7] &[30.6, 90.4]\\ 
    \subref{fig:f} &0.857 &\cmark &[10.6, 78.7] &[10.9, 78.3] &[22.0, 70.0]\\ 
    \midrule
    \rowcolor{black!3} 
    \subref{fig:g} &0.934 &\cmark &[19.1, 93.3] &[19.5, 90.4] &[22.0, 25.5]\\ 
    \subref{fig:h} &0.895 &\cmark &[19.1, 116 ] &[19.5, 110 ] &[30.5, 104 ]\\ 
    \rowcolor{black!3} 
    \subref{fig:i} &0.943 &\cmark &[10.6, 92.9] &[10.9, 84.3] &[21.9, 25.4]\\
    \bottomrule
  \end{tabularx}
\end{table}

When the capacitor size increases, the probability of failure paths decrease for all variants, which is expected. Configurations~\subref{fig:d},~\subref{fig:f},~\subref{fig:g},~\subref{fig:h}, and~\subref{fig:i} in Figure~\ref{fig:output-intermittent-execution} satisfy the timing requirement (see Line 3 in Figure~\ref{fig:example}) with 0.8 probability ($P_r(\tau_{classify} \le 40ms) > 0.8$). If we increase the capacitor size from 2mF to 5mF, the probability of meeting the requirement becomes 0.9 for all variants. This increase our confidence in meeting the requirement, especially for configurations~\subref{fig:g},~\subref{fig:h}, and~\subref{fig:i}. Program variants 1 and 3 have more stable timing behavior compared to variant 2 (less divergent timing behavior in the probability distributions), and variant 3 performs slightly better than variant 1 for all confidence intervals. Therefore, we can conclude that it would be better to use 5mF capacitor powering the program variant 3 for function \textit{classify} (see~\subref{fig:f}). 5mF capacitor has a higher energy level, and the probability of meeting the timing requirement would be higher due to less power failure probability. However, bigger capacitors would require more time to store enough energy to power up the application, which would decrease the sampling rate of the sensing application. In this analysis, the setup harvests energy from the emitted radio frequency (RF) signals at a distance of 40cm. The charging time distributions of the capacitors we used are given in 95\% confidence intervals as follows:
\begin{center}
  \begin{tabularx}{.75\linewidth}{@{}Xccc@{}}
    \toprule
    40\textit{cm} distance &\multicolumn{1}{c}{1\emph{mF}} &\multicolumn{1}{c}{2\emph{mF}} &\multicolumn{1}{c}{5\emph{mF}} \\
    \midrule
    {\small$\tau_{harvest}$} &[9.48\emph{ms}, 11.95\emph{ms}] &[18.99\emph{ms}, 25.64\emph{ms}] &[41.47\emph{ms}, 57.53\emph{ms}]\\ 
    \bottomrule
  \end{tabularx}
\end{center}
We can conclude that it would be better to \textit{minimize} the capacitor size while meeting the \textit{timing requirement}. Also, a 2mF capacitor would be the optimal selection for this setup under the given timing requirement and energy profile. 

\section{Evaluation}
\label{sec:evaluation}

We now evaluate our prototype to demonstrate that: (i) \sysname correctly predicts the execution time of intermittent programs and (ii) it significantly reduces the analysis time and efforts. 

\subsection{Testbed Setup}
\label{sec:evaluation-setup}

\subsubsection*{Target Platform.} We used TI's MSP430FR5994 LaunchPad~\cite{MSP430FR5994} development board as a target platform. The operating frequency of the microcontroller (MCU) was set to 1 MHz for our testbed experiments, however, we also investigated 4MHz, 8MHz, and 16MHz clock-rates. The MCU supports 20-bit registers and 20-bit addresses to access an address space of 1 MB. To infer instruction-level energy consumption and obtain timing models, we used TI's EnergyTrace software~\cite{EnergyTrace} sampling the energy consumption of programs at runtime by using the specialized debugger circuitry on the development board. The precision of EnergyTrace was limited due to its low sampling rate. 

\subsubsection*{Checkpoint Implementation.} This MCU has 256 KB of FRAM (non-volatile memory) to store data that persists when there is no power. It also includes 4 KB of SRAM (volatile memory) to store program variables with automatic scope. We implemented (i) a checkpoint routine (\emph{checkpoint()}) that copies the volatile computation state (20-bit general-purpose registers, program counter, and 4KB SRAM) into FRAM and (ii) a recovery routine (\emph{recovery()}) that restores the computation state after a power failure by using the latest successful checkpoint data. The energy and timing costs of these routines are constant. We configured \sysname with these costs by using pragmas, as shown in Figure~\ref{fig:example}.  

\subsubsection*{Energy Harvesting Tools.} We used the Powercast TX91501- 3W power transmitter~\cite{powercast} emitting radio frequency (RF) signals at 915 MHz center frequency. The transmitter was connected to a P2110-EVB receiver~\cite{P2110-EVB} co-supplied with a 6.1 dBi patch antenna. The receiver harvested energy from the emitted RF signals to power the MSP430FR5994 launchpad board. We used Arduino Uno~\cite{Arduino} with 10-bit ADC (analog to digital converter) to measure the instantaneous harvested power by P2110-EVB receiver.

\subsubsection*{Host Platform.} We run \sysname on an Intel Core i7-9750H CPU @ 2.60GHz $\times$ 12 core machine with 32 GiB RAM. \sysname used one core and 16 GiB memory on average. The memory consumption was relatively high, mainly due to the symbolic computation introduced by our benchmarks.

\begin{table}
\centering
  \caption{Stochastic Cost Models of the MSP430 Instruction Set Architecture based on Formats and Addressing Modes. 
  }
  \label{tab:target-modeling}
  \begin{tabularx}{.7\columnwidth}{@{}XcXX@{}} 
  \toprule
\multicolumn{1}{c}{Addressing}   &\multicolumn{1}{c}{Example} &\multicolumn{1}{c}{Timing} &\multicolumn{1}{c}{Energy}\\
{Modes} &{Instructions} &{Models ($\mu s$)} &{Models ($nj$)}\\
  \midrule
{Format I} &{Double Operand}\\
  \midrule
  \rowcolor{black!3} 
00 0 &\emph{ mov r5, r9 }         &$\normal$(1.02, 0.01)  &$\normal$(4.52, 0.62) \\
00 1 &\emph{ add r5, 3(r9)}       &$\normal$(3.02, 0.01)  &$\normal$(7.08, 0.62) \\
   \rowcolor{black!3} 
01 0 &\emph{ mov 2(r5),r7}        &$\normal$(3.02, 0.01)  &$\normal$(6.97, 0.62) \\
01 1 &\emph{ add 3(r4), 6(r9)}    &$\normal$(5.02, 0.01)  &$\normal$(10.1, 0.62) \\
   \rowcolor{black!3} 
10 0 &\emph{ and @r4, r5}         &$\normal$(2.02, 0.01)  &$\normal$(5.80, 0.62) \\
10 1 &\emph{ xor @r5, 8(r6)}      &$\normal$(4.02, 0.01)  &$\normal$(8.33, 0.62) \\
   \rowcolor{black!3} 
11 0 &\emph{ mov \#20, r9}        &$\normal$(2.02, 0.01)  &$\normal$(5.55, 0.62) \\
11 1 &\emph{ mov @r9+, 2(r4)}     &$\normal$(4.02, 0.01)  &$\normal$(8.34, 0.62) \\
\midrule
{Format II} &{Single Operand}\\
\midrule
   \rowcolor{black!3} 
{00\:\:\:} &\emph{push r5}        &$\normal$(3.01, 0.01)  &$\normal$(8.34, 0.62) \\
{01\:\:\:} &\emph{call 2(r7)}     &$\normal$(4.02, 0.01)  &$\normal$(10.1, 0.62) \\
   \rowcolor{black!3} 
{10\:\:\:} &\emph{push @r9}       &$\normal$(3.52, 0.01)  &$\normal$(8.33, 0.62) \\
{11\:\:\:} &\emph{call \#81h}     &$\normal$(4.02, 0.01)  &$\normal$(10.1, 0.62) \\
  \midrule
{Format III} &{Special Instructions}\\
  \midrule 
  \rowcolor{black!3} 
Jxx &\emph{ jmp r5, r9 } &\emph{Constant}(2)  &$\normal$(5.8, 0.62) \\
mult. &\emph{call $\#\_$mspabi$\_$mpyi} &$\normal$(15.94, 0.27)  &$\normal$(16.38, 0.23) \\   
  \rowcolor{black!3} 
div.  &\emph{call $\#\_$mspabi$\_$divu} &$\normal$(16.39, 0.23)  &$\normal$(16.68, 0.17) \\
\bottomrule
  \end{tabularx}
\end{table}

\subsection{Profiling Target Platform and Energy Environment}

The target platform and energy environment profiling is a one-time effort and eliminates the need for continuous target deployment and on-the-fly analysis efforts. 

\subsubsection*{Profiling Target Platform.}

\sysname runs on the \emph{platform-independent} LLVM instruction set~\cite{LLVM_IR}. There is no one-to-one correspondence between the LLVM instruction set 
and the target architecture instruction set. Therefore, we employed a block-based mapping strategy as mentioned in Section~\ref{sec:cost_models}. In the cost model generation step of \sysname (Step 1 in Figure~\ref{fig:approach}), the target hardware instructions were automatically mapped into the LLVM basic blocks. The time and energy cost distributions of each basic block were calculated through the hardware instruction cost distributions.

We used the Saleae logic analyzer~\cite{Saleae} and EnergyTrace tool to collect the energy and timing costs of the MSP430 instructions from the MSP430FR5994 Launchpad at 1MHz clock frequency. Table~\ref{tab:target-modeling} presents a simplified version of the cost model we used to predict basic block costs. MSP430 is a 16-bit RISC instruction-set architecture with no data-cache~\cite{epic}. It has 27 native and 24 emulated instructions. Our analysis shows that the energy and timing costs of the instructions depend on the formats and addressing modes. The MSP430 architecture has seven modes to address its operands: register (00), indexed (01), absolute (10), immediate (11), symbolic, register indirect, and register indirect auto increment. The timing and energy behavior of each instruction highly depends on the these addressing modes, which we can group as double operand instructions (Format I), single operand instructions (Format II), and special instructions (Format III)~\cite{msp430datasheet}. We sampled the addressing modes and their combinations grouped by their formats, and then statistically inferred the distribution of sample means employing bootstrapping method~\cite{lock2020statistics}. 

Apart from those instructions, we modeled the subroutines, \emph{call $\#$mspabi$\_$mpyi} and \emph{call $\#$mspabi$\_$divu}, generated by the compiler since MSP430 does not have a hardware multiplier and divider. Besides, we derived regression models for \textit{call $\#$memset} and \textit{call $\#$memcopy} intrinsic functions. For \textit{call $\#$memcopy}, the stochastic models for timing and energy are $\#\text{memcpy}_{\tau} \sim \normal(13.06, .01) \ast c \cdot \normal(9.06,.01)$ and $\#\text{memcpy}_{\epsilon} \sim \normal(35.04, 1.38) \ast c \cdot \normal(24.21,1.24)$ where $c$ is the explanatory variable for the number of data words copied.

\begin{figure}
    \centering
\centerline{\includegraphics[width=.8\columnwidth]{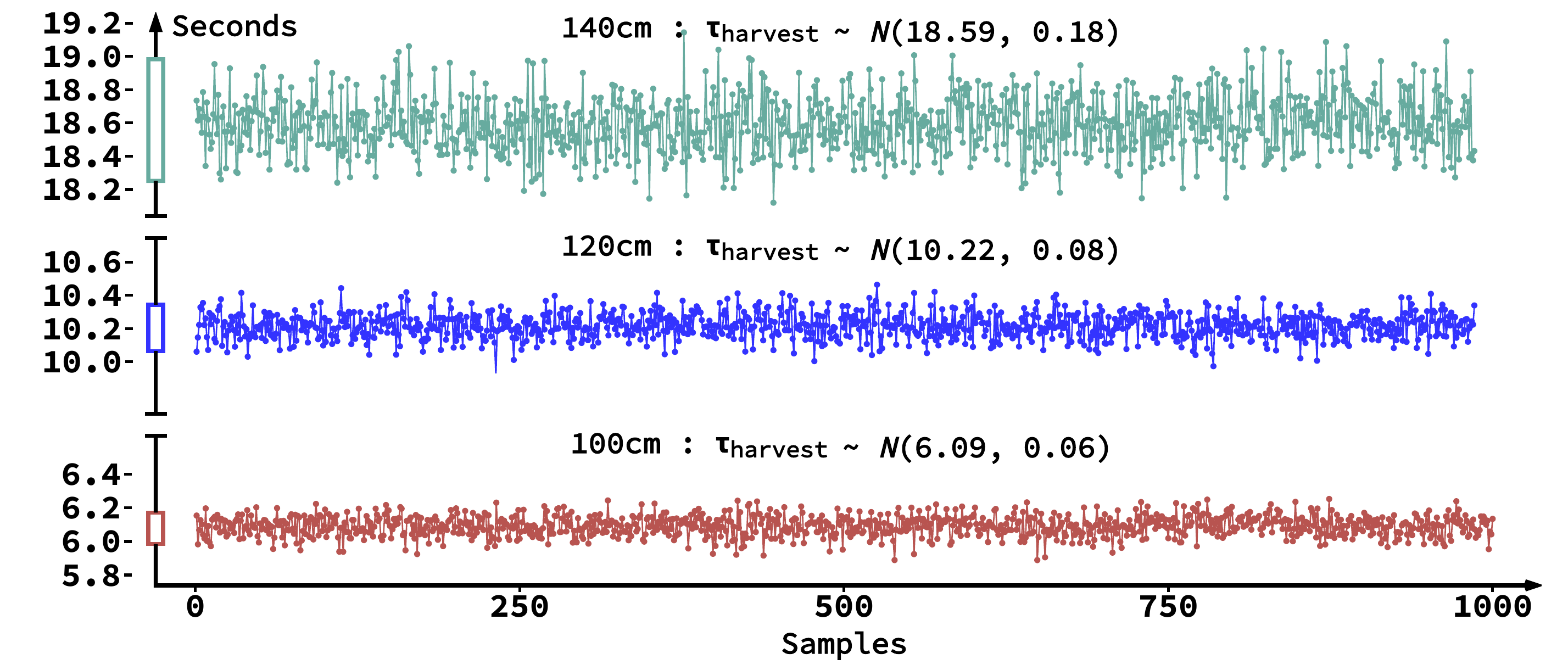}}
    \caption{RF energy profiles: charging-time samples for a 50\emph{mF} supercapacitor at different distances.} 
    \label{fig:energy-profile}
\end{figure}

\subsubsection*{Energy Profiles of the Environment}

\sysname requires the ambient energy profile and the capacitor size to infer power failure rates and the off- and on-times of the device. Therefore, we collected the ADC samples of the instantaneously received power from different setups for 20 seconds. We created different settings by placing the RF power transmitter and receiver in line of sight at different distances. The measurements were repeated ten times for each setup. Using the energy profile and energy equation of capacitors ($E = \nicefrac{1}{2} \cdot C \cdot V^2$), one can infer a bootstrapping distribution~\cite{lock2020statistics} to model the average waiting time required to charge a capacitor with the given voltage threshold.  We run our benchmarks using energy harvesting hardware. Since the instantaneous power obtained from energy harvesting devices varies depending on the distance between the receiver and the transmitter, 
we chose three different distances in our testbed (100\emph{cm}, 120\emph{cm}, and 140\emph{cm}). Figure~\ref{fig:energy-profile} presents the energy profiles sampled at these distances. The MCU starts to operate \textit{intermittently} at 100\emph{cm} to the transmitter. The harvested power significantly decreases after 120\emph{cm}, and the MCU dies quite frequently.

\subsection{Evaluation Results}

\begin{table}
\centering
  \caption{Benchmarks. Instructions and block counts are derived while MSP430 is \emph{continuously-powered}.}
  \label{tab:benchmarks-explained}
  \begin{tabularx}{.7\linewidth}{@{}lXXXXcX@{}}
    \toprule
    Benchmark & \multicolumn{2}{c}{\# of Instructions} & \multicolumn{2}{c}{\# of Blocks} &\multicolumn{1}{c}{\# of Power} & Source\\
    ~ & \emph{Static} & \emph{Dynamic} & \emph{Static} & \emph{Dynamic} & Failure Paths\\
    \midrule
    \rowcolor{black!3} 
    BitCount   &566 &103702 &66 &7439 & 922 & \cite{mibench}\\
    CRC        &92  &1568 &17 &408 & 363 & \cite{mibench}\\
    \rowcolor{black!3} 
    Dijkstra   &282 &24585 &36 &2884 & 592 & \cite{mibench}\\
    FIR Filter &390 &144400 &13 &1880 & 175 & \cite{embench}\\
    \bottomrule
  \end{tabularx}
\end{table}

We used four benchmarks having different computation demands and memory access frequency affecting the execution time (see Table~\ref{tab:benchmarks-explained}). 
In particular, our benchmarks implement network algorithms and data filtering in real IoT and Edge computing applications. 



\begin{table}
  \centering
  \caption{\textbf{Prediction for Intermittent Execution Times} compared with Actual Measurements ($\mu$--mean, $\sigma$--standard deviation). 
  }

  \label{tab:intermittent-execution-results}
  \begin{tabularx}{.7\columnwidth}{@{}l|X|XX|XX|X}
    \toprule
    \multicolumn{1}{l|}{Benchmark} &\multicolumn{1}{c|}{Distance} &\multicolumn{2}{c|}{Actual Measure. ($m$s)} &\multicolumn{2}{c|}{\sysname Prediction ($m$s)} &\multicolumn{1}{X}{\cellcolor{babyblueeyes!25} \bf Error}\\
    \multicolumn{1}{c|}{~} &\multicolumn{1}{c|}{(cm)} &\multicolumn{1}{c}{$\mu$} &\multicolumn{1}{c|}{$\sigma$} &\multicolumn{1}{c}{$\mu$} &\multicolumn{1}{c|}{$\sigma$} &\multicolumn{1}{c}{\cellcolor{babyblueeyes!25} \bf in \%}\\
\midrule
    \rowcolor{black!3} 
BitCount     &\emph{<100}&6150.038   &0.013    &6145.405   &0.003   &\cellcolor{babyblueeyes!25}\enspace 0.075\\
{ }          &\emph{100} &33550.353  &19.919   &33693.902  &99.793  &\cellcolor{babyblueeyes!25} 0.4279\\
\rowcolor{black!3} 
{ }          &\emph{120} &99815.832  &1008.207 &99618.170  &303.237 &\cellcolor{babyblueeyes!25} 0.0197\\
{ }          &\emph{140} &206575.550 &219.747  &206546.729 &267.430 &\cellcolor{babyblueeyes!25} 0.0143\\
\rowcolor{black!3} 
\midrule
CRC          &\emph{<100}&3689.654   &0.001    &3690.537   &0.001   &\cellcolor{babyblueeyes!25}\enspace 0.022\\
{ }          &\emph{100} &23337.998  &23.339   &23655.506  &29.708  &\cellcolor{babyblueeyes!25} 1.3605\\
\rowcolor{black!3} 
{ }          &\emph{120} &54626.126  &771.290  &54048.093  &154.173 &\cellcolor{babyblueeyes!25} 1.0582\\
{ }          &\emph{140} &95888.740  &1369.499 &96297.124  &421.776 &\cellcolor{babyblueeyes!25} 0.4259\\
\rowcolor{black!3} 
\midrule
Dijkstra     &\emph{<100}&3212.243   &0.013    &3201.285   &0.006   &\cellcolor{babyblueeyes!25}\enspace  0.341\\
{ }          &\emph{100} &27136.569  &459.938  &27469.100  &56.679  &\cellcolor{babyblueeyes!25} 1.2254\\
\rowcolor{black!3} 
{ }          &\emph{120} &40455.846  &636.112  &40506.445  &74.279  &\cellcolor{babyblueeyes!25} 0.1251\\
{ }          &\emph{140} &73032.293  &88.785   &73026.767  &156.892 &\cellcolor{babyblueeyes!25} 0.0076\\
\rowcolor{black!3} 
\midrule
FIR Filter   &\emph{<100}&7282.946   &0.019    &7277.968   &0.004   &\cellcolor{babyblueeyes!25}\enspace  0.068\\
{ }          &\emph{100} &51716.969  &33.936   &51664.758  &173.920 &\cellcolor{babyblueeyes!25} 0.1010\\
\rowcolor{black!3} 
{ }          &\emph{120} &97558.188  &98.225   &97702.685  &136.667 &\cellcolor{babyblueeyes!25} 0.1481\\
{ }          &\emph{140} &163037.556 &138.595  &163634.595 &481.34  &\cellcolor{babyblueeyes!25} 0.3662\\
\bottomrule
\end{tabularx}

\end{table}

\subsubsection*{\sysname Prediction Accuracy.}

We compared \sysname's predictions to the actual measurements obtained from the testbed (see Section~\ref{sec:evaluation-setup} for the evaluation setup). The difference between the \sysname estimation and the actual measurements varies between 0.007\% and 1.3\% (see Table~\ref{tab:intermittent-execution-results}). Despite the limitations of our measurement devices, and in turn, coarse-grained ambient energy profiles, \sysname predicted the power failure probabilities with high accuracy. Since the device starts to operate continously at a distance of less than 100cm, the prediction error significantly decreases to at most 0.34\% and at least 0.022\%. Moreover, it accurately modeled the charging and discharging times by using the probabilistic distribution of the ambient energy. Note that the prediction accuracy of \sysname depends on the accuracy of the hardware cost models, and the ambient energy profile. 

The prediction errors in Table~\ref{tab:intermittent-execution-results} (e.g., \sysname 's predictions for CRC and Dijkstra are less accurate at closer distances to the energy harvester) are due to our imperfect probabilistic models. Moreover, \sysname does not consider the fact that the battery-free device is simultaneously charging while discharging. Therefore, as the distance gets longer, the amount of charge loaded on the capacitor during the discharge period decreases, and the predictions become more accurate.

\begin{tcolorbox}[boxsep=5pt,left=2pt,right=2pt,top=2pt,bottom=2pt]
\paragraph{\textbf{Summary.}} \sysname predicted the execution time of RF-powered intermittent programs \textbf{almost perfectly}. We observed a reasonable maximum prediction error during our evaluation, which was less than 1.5\%.
\end{tcolorbox}

\subsubsection*{\sysname Analysis Time.}

\begin{table}[t]
	\centering
	\caption{\textbf{Time Study} for Intermittent Execution. \sysname's automated analysis compared to manual experiment times.}
	\label{table:time-study}
	\begin{tabularx}{.75\linewidth}{l|c|c|XXX|X} 
		\toprule
		Benchmark &\multicolumn{1}{c|}{\sysname Analysis} &\multicolumn{1}{c|}{Harv. Profile}& \multicolumn{3}{c|}{Experiment Time (s)} & \multicolumn{1}{c}{Cap.}\\
		~ & {Time (s)} & {Time (s)} & \emph{100 cm } & \emph{120 cm} & \emph{140 cm} & {size} \\
		\midrule
    \rowcolor{black!3} 
		Bitcount   &54   &200 &1698 &5042 &10412 &50mF\\
		CRC        &5    &200 &1249 &2797 &3895  &50mF\\
    \rowcolor{black!3} 
		Dijkstra   &24   &200 &1391 &2029 &3842  &50mF\\
		Fir Filter &26   &200 &2591 &4897 &8203  &50mF\\
		\bottomrule
	\end{tabularx}
\end{table}

To assess how significantly \sysname reduces the analysis time and efforts, we measured the time required to deploy an RF-powered application onto our testbed (see Section~\ref{sec:evaluation-setup}), run the application 50 times, and collect a sufficient amount of data to reason about the timing behavior through statistical sampling. The 4\textsuperscript{th} column in Table~\ref{table:time-study} presents the time it takes for the experimental measurements at each distance and excludes the manual analysis efforts such as data extraction and preparation for data analysis. 

As shown in Table~\ref{table:time-study}, the manual experiment time increases as the number of power failures during intermittent operation increases (as the distance from the transmitter increases) while \sysname's analysis time does not change with regard to distances of the harvesting kit, and it mainly depends on the number of dynamic program blocks and instruction counts (see Table~\ref{tab:benchmarks-explained}). Moreover, \sysname checks the likelihood of a power failure after each block, which increases its analysis time. \sysname appreantly requires less manual effort and is significantly faster than manual testing. \sysname performed more precise path-based symbolic analysis 
at least 100 times faster than experimental measurements under low energy harvesting conditions.

\begin{tcolorbox}[boxsep=5pt,left=2pt,right=2pt,top=2pt,bottom=2pt]
\paragraph{\textbf{Summary.}} \sysname speeds up the timing analysis of intermittent programs by \textbf{at least two orders of magnitude} and eliminates the burden of the manual data analysis effort. 
\end{tcolorbox}

\section{Related Work}
\label{sec:related}

We present prior work that relates to our approach in the context of intermittent computing, timing and energy analysis of intermittent programs, and probabilistic program analysis.

\subsubsection*{Timing Analysis of Intermittent Programs.} Some runtimes provide programming constructs to assign timestamps to sensed data and check if the data expire. InK~\cite{ink} is a reactive task-based runtime that employs preemptive and power failure-immune scheduling for timing constraints of task threads. Mayfly~\cite{mayfly} makes the passing of time explicit, binding data to the time it was collected, and keeping track of data and time through power failures. TICS~\cite{tics} provides programming abstractions for handling the passing of time through intermittent failures and making decisions about when data can be used or thrown away. Different from these runtimes, \sysname predicts the timing behavior of intermittent programs before deployment and introduces zero overhead.

\subsubsection*{Energy Analysis of Intermittent Programs.} CleanCut~\cite{cleancut} detects non-terminating tasks in a task-based intermittent program. To do so, it samples the energy consumption of program blocks with a special debugging hardware and over approximates path-based energy consumption. Similar to CleanCut, EPIC~\cite{epic} traverses control-flow graph to compute best- and worst-case estimates of dynamic energy consumption for a given intermittent program. \sysname similarly analyze the energy costs of paths but considers precise program semantics and knows execution probabilities of each path by means of probabilistic symbolic execution. In fact, profiling is a one-time effort for each target and has intrinsic support for detection of forward-progress violations. \sysname's main goal is to perform energy-aware timing analysis to reason about timing behavior of intermittent programs. 

\subsubsection*{Other Analysis Tools for Intermittent Programs.} IBIS~\cite{ibis} performs a static taint analysis to detect bugs caused by non-idempotent I/O operations in intermittent systems. ScEpTIC~\cite{sceptic} has similar objectives as IBIS, i.e., detecting intermittent bugs. Ocelot~\cite{ocelot} enforces intermittent programs written rust language to maintain temporal consistency. Those tools are complementary and can be used with \sysname to detect such intermittence bugs. But none of these tools perform timing analysis of intermittent programs. Moreover, \sysname is a novel probabilistic program analysis approach.

\subsubsection*{Probabilistic Program Analysis.}

Various probabilistic symbolic execution techniques have been proposed in the literature (e.g.,~\cite{geldenhuys2012probabilistic, dwyer2015probabilistic, puasuareanu2010symbolic,luckow2014exact, borges2015iterative, chen2016generating, filieri2013reliability}). Geldenhuys et al.~\cite{geldenhuys2012probabilistic} propose probabilistic symbolic execution as an extension of Symbolic PathFinder~\cite{puasuareanu2010symbolic}. 
Luckow et al.~\cite{luckow2014exact} extend probabilistic symbolic execution
to compute a scheduler resolving program non-determinism to maximize the property satisfaction probability.
Chen et al.~\cite{chen2016generating} employ probabilistic symbolic execution to generate a performance distribution that captures the input probability distribution over the execution times of the program. Filieri et al.~\cite{filieri2013reliability} extract failure and success program paths to be used for probabilistic reliability assessment against constraints representing subsets of all input variables range over finite discrete domains. All these techniques consider only typical program non-determinism, e.g., program input probability distribution. They analyze the control flow of continuously powered program execution, and they do not consider the power-failure-induced control flow. To analyze the power-failure-induced control flow, we need symbolic execution at the IR- (e.g., LLVM and VEX) or at the binary level, which is not supported by the current probabilistic symbolic execution engines. Therefore, it is not feasible to exploit, integrate or reuse the existing techniques for the analysis of intermittent programs. Our goal is to overcome this problem and develop a dedicated probabilistic symbolic execution technique that analyzes the control flow graph of the intermittent programs. That’s why we introduce power-failure-induced edges, and we use the charging/discharging model, environment energy profile, and other intermittent program characteristics (capacitor size and program structure). 

\section{Discussion}
\label{sec:discussion}

\subsubsection*{Comparison to simple stochastic simulation.} Stochastic simulations may not cover all execution paths, since they randomly generate simulation inputs and simulate the program based on these input values. In order to execute all possible paths, we would need to perform a large number of simulation runs. Even then, there might still be some paths that have not been examined, and it is not always possible to give a good estimate how many simulations we need to obtain a result accurate enough. Contrarily, \sysname employs probabilistic symbolic execution to get the absolute path frequency by using probability distributions, rather than randomly generating inputs and executing the program paths. This is a more accurate approach compared to simulation, since we are guaranteed to cover all execution paths needed to analyze the timing behavior of the intermittent program.

\subsubsection*{Scalability of probabilistic symbolic execution.} We do not expect serious scalability issues for \sysname analyzing intermittent programs running on batterless devices since these programs are relatively small by nature. This is also why we chose our benchmarks among the most common benchmarks that are widely accepted by the intermittent computing community (e.g.,~\cite{chain, cleancut}). We already presented the analysis times for the four applications in Table~\ref{table:time-study}’s \sysname Analysis Time column. The Bitcount benchmark took about a minute. We also believe that a thorough scalability analysis will be beneficial to understand the limits of \sysname in terms of the maximum program size that we can analyze as of now.

\subsubsection*{The prediction accuracy achieved by \sysname in the unpredictable nature of energy harvesting.} The environment models that have more inherent variability will influence the accuracy and precision of the prediction negatively such as energy modeling with mobile transmitter or receiver or both. But the focus of \sysname is not on generating the environment model, but on performing the analysis given user-provided model. The user can provide a more sophisticated probabilistic model for incoming profiles, bimodal, mixture, etc. On the other hand, the RF profile is not stable, it is actually chaotic. But it can be modeled with a Gaussian distribution easily at certain distances (by applying bootstrapping method on energy traces) as long as the testbed and the RF source are not mobile. We model the environmental energy profile in terms of the average waiting time required to fully charge a capacitor with specific capacitance values on the harvester kit. For instance, the average waiting time required to charge a 50mF capacitor varies from 8 seconds to 9.5 seconds at 100 cm away from the RF source. And the more the testbed is positioned away from the source, the waiting time exponentially increases.

\subsubsection*{Dealing with loops.} If the program has a bounded-loop, simply unrolling the loop would be sufficient; however, there might be conditions data-dependent upon the program’s input 
that result in branch points in the computation tree. In these cases, the path exploration algorithm (Algorithm~\ref{algo:dfs}) traverses the loop until the probability of branching approaches zero. Our computation tree might have infinite depth, as some loops may be unbounded. Therefore, \sysname terminates analysis after exploring a user-provided limit (Line 6 in Algorithm~\ref{algo:dfs}). In addition, all the benchmarks used in the paper include loops.

\subsubsection*{Timing requirements.} Timing requirements are optional inputs of \sysname. Without timing requirements, \sysname reports the timing distribution of each function in the LLVM module. Programmers only provide the function to be analyzed, and \sysname generates distributions for the timing and energy consumption of that function. If a more fine-grained analysis is needed, timing requirements can be input to get a quantification report about the success rate of the requirements.

\subsubsection*{Ambient energy profile.} In designing \sysname, we assume that programmers follow a what-if analysis and evaluate checkpoint placement and timing behavior of functions under different ambient energy profiles. For instance, three profiles can be sufficient to make judgments for the solar energy harvesting case: 
\textit{high}, \textit{medium}, and \textit{low}, representing environmental effects that gradually vary from sunny to cloudy conditions. In rare cases, it would be useful to provide a perfect ambient energy model.

\subsubsection*{Energy cost model.} Our evaluation shows that our energy cost model is already sufficient to perform an accurate timing analysis of intermittent programs. We observed less than 2\% estimation error with our approach considering the benchmarked applications. Therefore, we did not see any reason to devise or incorporate a better model. It is possible to integrate better models (i.e., more accurate probability distributions that can provide a fine-grained representation of the pipelining and cache effects) into \sysname to increase the analysis accuracy. Considering the focus of our paper (i.e., probabilistic symbolic execution approach and intermittent computing) proposing a better energy cost model is orthogonal to our contributions.

\subsubsection*{Requirement of hardware support and cost of energy profiling.} \sysname can use probabilistic or analytical models derived from either the real measurements conducted on the hardware, or directly provided (for example, by a hardware vendor). There are many open-source energy harvesting and power consumption traces available, e.g.,~\cite{enhants}. Therefore, \sysname users can analyze their programs without the need for any hardware measurements, as long as they are provided via some of the mentioned means.

\subsubsection*{Using energy approximation.} One may argue that compile-time analysis using energy cost approximations (e.g.,~\cite{baghsorkhi2018automating}) can be employed to predict the execution time of intermittent programs. Approximating the energy consumption of a code block can only give the time it takes to execute it continuously. The code block can be interrupted at any point during its execution, and there are extra recovery operations due to power failures. This situation leads to power-failure-induced control flow, which is highly dynamic. Therefore, it is impossible to infer the execution time by using simple energy consumption approximations. And, we need a custom technique to infer the execution time concerning the dynamic power-failure induced control flow.

\subsubsection*{Drift between distributions.} There might be a drift between the distributions at measurement versus deployment time. It may not always be possible to know the exact energy harvesting and sensor input distributions in the field. Therefore, we design \sysname to give programmers insight into how their intermittent programs behave under different ambient energy and sensor input conditions. Programmers are expected to use \sysname with different sensor input and energy harvesting distributions and estimate the timing behavior of their programs. 

\subsubsection*{MSP430.} As MSP430 is a de facto standard for intermittent computing, we target the MSP430 instruction set and its implementation, the MSP430FR5994 board, to assess \sysname. We empirically show that the energy and timing behavior of the instruction set could be modeled through sampling distributions. In the future, we will explore more sophisticated chips.

\section{Conclusion}
\label{sec:conclusion}

We presented a novel static analysis approach, \sysname, that estimates the timing behavior of intermittent programs, affected by several factors such as ambient energy, the power consumption
of the target hardware, capacitor size, program input space,
and program structure. 
Considering the effects of power failures, \sysname symbolically executes the given program to generate the execution time probability distributions of each
function in the program. To do so, it requires probabilistic energy and timing cost models of the target platform, capacitor size, and program input space. 
Our evaluation 
showed that \sysname exhibits a worst-case prediction accuracy of 99.6\% for a set of benchmark codes. 

Using the output of \sysname, programmers can follow a what-if analysis, e.g., reconfiguring the hardware and/or restructuring their program to ensure the desired timing behavior of their program. This what-if analysis is currently manual and not guided. As future work, we plan to employ metaheuristic algorithms \cite{goldberg1988genetic,van1987simulated} to guide the analysis \cite{harman2001search,HarmanMSY10}.


\section*{Acknowledgments} This work was partially supported by the SINTEF project G-IoT (Green Internet of Things), funded by the basic funding through Research Council of Norway.

\bibliographystyle{plain}
\bibliography{references}

\begin{thebibliography}{10}

\bibitem{epic}
Saad Ahmed, Muhammad Nawaz, Abu Bakar, Naveed~Anwar Bhatti, Muhammad~Hamad
  Alizai, Junaid~Haroon Siddiqui, and Luca Mottola.
\newblock Demystifying energy consumption dynamics in transiently powered
  computers.
\newblock {\em ACM Transactions on Embedded Computing Systems (TECS)},
  19(6):1--25, 2020.

\bibitem{embench}
Mario Almeida, Stefanos Laskaridis, Ilias Leontiadis, Stylianos~I. Venieris,
  and Nicholas~D. Lane.
\newblock Embench: Quantifying performance variations of deep neural networks
  across modern commodity devices.
\newblock In {\em the 3rd International Workshop on Deep Learning for Mobile
  Systems and Applications (EMDL'19)}, page 1–6, 2019.

\bibitem{Arduino}
{Arduino}.
\newblock Arduino uno rev-3.
\newblock \url{https://store.arduino.cc/usa/arduino-uno-rev3}, 2021.
\newblock Last accessed: Aug. 10, 2021.

\bibitem{baghsorkhi2018automating}
Sara~S Baghsorkhi and Christos Margiolas.
\newblock Automating efficient variable-grained resiliency for low-power iot
  systems.
\newblock In {\em Proceedings of the 2018 International Symposium on Code
  Generation and Optimization (CGO'18)}, pages 38--49, 2018.

\bibitem{symexsurvey}
Roberto Baldoni, Emilio Coppa, Daniele~Cono D’elia, Camil Demetrescu, and
  Irene Finocchi.
\newblock A survey of symbolic execution techniques.
\newblock {\em ACM Computing Surveys (CSUR)}, 51(3):1--39, 2018.

\bibitem{hibernus++}
Domenico Balsamo, Alex~S Weddell, Anup Das, Alberto~Rodriguez Arreola, Davide
  Brunelli, Bashir~M Al-Hashimi, Geoff~V Merrett, and Luca Benini.
\newblock Hibernus++: a self-calibrating and adaptive system for
  transiently-powered embedded devices.
\newblock {\em IEEE Transactions on Computer-Aided Design of Integrated
  Circuits and Systems}, 35(12):1968--1980, 2016.

\bibitem{harvos}
Naveed~Anwar Bhatti and Luca Mottola.
\newblock Harvos: Efficient code instrumentation for transiently-powered
  embedded sensing.
\newblock In {\em the 16th ACM/IEEE International Conference on Information
  Processing in Sensor Networks (IPSN'17)}, pages 209--220, 2017.

\bibitem{blitzstein2019introduction}
Joseph~K Blitzstein and Jessica Hwang.
\newblock {\em Introduction to probability}.
\newblock Crc Press, 2019.

\bibitem{borges2015iterative}
Mateus Borges, Antonio Filieri, Marcelo d'Amorim, and Corina~S
  P{\u{a}}s{\u{a}}reanu.
\newblock Iterative distribution-aware sampling for probabilistic symbolic
  execution.
\newblock In {\em the 10th Joint Meeting on Foundations of Software Engineering
  (ESEC/FSE'15)}, pages 866--877, 2015.

\bibitem{klee}
Cristian Cadar and Martin Nowack.
\newblock Klee symbolic execution engine in 2019.
\newblock {\em International Journal on Software Tools for Technology
  Transfer}, pages 1--4, 2020.

\bibitem{chen2016generating}
Bihuan Chen, Yang Liu, and Wei Le.
\newblock Generating performance distributions via probabilistic symbolic
  execution.
\newblock In {\em the 38th International Conference on Software Engineering
  (ICSE'16)}, pages 49--60, 2016.

\bibitem{Clang}
{Clang}.
\newblock {Clang}: a c language family frontend for {LLVM}.
\newblock \url{https://clang.llvm.org/}, 2021.

\bibitem{chain}
Alexei Colin and Brandon Lucia.
\newblock Chain: tasks and channels for reliable intermittent programs.
\newblock In {\em the 31st annual ACM Conference on Object-Oriented
  Programming, Systems, Languages and Applications (OOPSLA'16)}, pages
  514--530, 2016.

\bibitem{cleancut}
Alexei Colin and Brandon Lucia.
\newblock Termination checking and task decomposition for task-based
  intermittent programs.
\newblock In {\em the 27th International Conference on Compiler Construction
  (CC'18)}, pages 116--127, 2018.

\bibitem{z3}
Leonardo De~Moura and Nikolaj Bj{\o}rner.
\newblock Z3: An efficient smt solver.
\newblock In {\em International conference on Tools and Algorithms for the
  Construction and Analysis of Systems}, pages 337--340. Springer, 2008.

\bibitem{engage}
Jasper de~Winkel, Vito Kortbeek, Josiah Hester, and Przemys{\l}aw Pawe{\l}czak.
\newblock Battery-free game boy.
\newblock {\em Proceedings of the ACM on Interactive, Mobile, Wearable and
  Ubiquitous Technologies}, 4(3):1--34, 2020.

\bibitem{dwyer2015probabilistic}
Matthew~B Dwyer, Antonio Filieri, Jaco Geldenhuys, Mitchell Gerrard, Corina~S
  P{\u{a}}s{\u{a}}reanu, and Willem Visser.
\newblock Probabilistic program analysis.
\newblock In {\em International Summer School on Generative and
  Transformational Techniques in Software Engineering}, pages 1--25. Springer,
  2015.

\bibitem{filieri2013reliability}
Antonio Filieri, Corina~S P{\u{a}}s{\u{a}}reanu, and Willem Visser.
\newblock Reliability analysis in symbolic pathfinder.
\newblock In {\em the 35th International Conference on Software Engineering
  (ICSE'13)}, pages 622--631. IEEE, 2013.

\bibitem{preact}
Kai Geissdoerfer, Raja Jurdak, Brano Kusy, and Marco Zimmerling.
\newblock Getting more out of energy-harvesting systems: Energy management
  under time-varying utility with preact.
\newblock In {\em the 18th International Conference on Information Processing
  in Sensor Networks (IPSN'19)}, pages 109--120, 2019.

\bibitem{geldenhuys2012probabilistic}
Jaco Geldenhuys, Matthew~B Dwyer, and Willem Visser.
\newblock Probabilistic symbolic execution.
\newblock In {\em the 21st International Symposium on Software Testing and
  Analysis (ISSTA'12)}, pages 166--176, 2012.

\bibitem{Pragma}
{GNU}.
\newblock {The Pragma Directive}.
\newblock \url{https://gcc.gnu.org/onlinedocs/cpp/Pragmas.html}, 2021.

\bibitem{dnn}
Graham Gobieski, Brandon Lucia, and Nathan Beckmann.
\newblock Intelligence beyond the edge: Inference on intermittent embedded
  systems.
\newblock In {\em the 24th International Conference on Architectural Support
  for Programming Languages and Operating Systems (ASPLOS'19)}, pages 199--213,
  2019.

\bibitem{goldberg1988genetic}
David~E Goldberg and John~Henry Holland.
\newblock Genetic algorithms and machine learning.
\newblock {\em Machine Learning}, 1988.

\bibitem{gomes2009model}
Carla~P Gomes, Ashish Sabharwal, and Bart Selman.
\newblock Model counting.
\newblock In {\em Handbook of satisfiability}, pages 633--654. IOS press, 2009.

\bibitem{mibench}
Matthew~R Guthaus, Jeffrey~S Ringenberg, Dan Ernst, Todd~M Austin, Trevor
  Mudge, and Richard~B Brown.
\newblock Mibench: A free, commercially representative embedded benchmark
  suite.
\newblock In {\em the 4th annual IEEE international Workshop on Workload
  Characterization (WWC'01)}, pages 3--14. IEEE, 2001.

\bibitem{gutruf2018fully}
Philipp Gutruf, Vaishnavi Krishnamurthi, Abraham V{\'a}zquez-Guardado, Zhaoqian
  Xie, Anthony Banks, Chun-Ju Su, Yeshou Xu, Chad~R Haney, Emily~A Waters,
  Irawati Kandela, et~al.
\newblock Fully implantable optoelectronic systems for battery-free, multimodal
  operation in neuroscience research.
\newblock {\em Nature Electronics}, 1(12):652--660, 2018.

\bibitem{harman2001search}
Mark Harman and Bryan~F Jones.
\newblock Search-based software engineering.
\newblock {\em Information and software Technology}, 43(14):833--839, 2001.

\bibitem{HarmanMSY10}
Mark Harman, Phil McMinn, Jerffeson {Teixeira de Souza}, and Shin Yoo.
\newblock Search based software engineering: Techniques, taxonomy, tutorial.
\newblock In {\em Empirical Software Engineering and Verification -
  International Summer Schools, {LASER} 2008-2010, Revised Tutorial Lectures},
  volume 7007 of {\em Lecture Notes in Computer Science}, pages 1--59.
  Springer, 2010.

\bibitem{flicker}
Josiah Hester and Jacob Sorber.
\newblock Flicker: Rapid prototyping for the batteryless internet-of-things.
\newblock In {\em the 15th ACM Conference on Embedded Network Sensor Systems
  (SenSys'17)}, pages 19:1--19:13, 2017.

\bibitem{mayfly}
Josiah Hester, Kevin Storer, and Jacob Sorber.
\newblock Timely execution on intermittently powered batteryless sensors.
\newblock In {\em the 15th ACM Conference on Embedded Networked Sensor Systems
  (SenSys'17)}, pages 1--13, 2017.

\bibitem{hicks2017clank}
Matthew Hicks.
\newblock Clank: Architectural support for intermittent computation.
\newblock {\em ACM SIGARCH Computer Architecture News}, 45(2):228--240, 2017.

\bibitem{zygarde}
Bashima Islam and Shahriar Nirjon.
\newblock Zygarde: Time-sensitive on-device deep inference and adaptation on
  intermittently-powered systems.
\newblock {\em Proceedings of the ACM on Interactive, Mobile, Wearable and
  Ubiquitous Technologies}, 4(3):1--29, 2020.

\bibitem{quickrecall}
Hrishikesh Jayakumar, Arnab Raha, and Vijay Raghunathan.
\newblock Quickrecall: A low overhead hw/sw approach for enabling computations
  across power cycles in transiently powered computers.
\newblock In {\em the 27th International Conference on VLSI Design and 13th
  International Conference on Embedded Systems}, pages 330--335. IEEE, 2014.

\bibitem{konstantopoulos2015converting}
Christos Konstantopoulos, Eftichios Koutroulis, Nikolaos Mitianoudis, and
  Aggelos Bletsas.
\newblock Converting a plant to a battery and wireless sensor with scatter
  radio and ultra-low cost.
\newblock {\em IEEE Transactions on Instrumentation and Measurement},
  65(2):388--398, 2015.

\bibitem{tics}
Vito Kortbeek, Kasim~Sinan Yildirim, Abu Bakar, Jacob Sorber, Josiah Hester,
  and Przemys{\l}aw Pawe{\l}czak.
\newblock Time-sensitive intermittent computing meets legacy software.
\newblock In {\em the 25th International Conference on Architectural Support
  for Programming Languages and Operating Systems (ASPLOS'20)}, pages 85--99,
  2020.

\bibitem{lattner2004llvm}
Chris Lattner and Vikram Adve.
\newblock Llvm: A compilation framework for lifelong program analysis \&
  transformation.
\newblock In {\em the International Symposium on Code Generation and
  Optimization (CGO'04)}, pages 75--86. IEEE, 2004.

\bibitem{LLVM_IR}
{LLVM}.
\newblock {LLVM} language reference manual.
\newblock \url{https://llvm.org/docs/LangRef.html}, 2021.

\bibitem{lock2020statistics}
Robin~H Lock, Patti~Frazer Lock, Kari~Lock Morgan, Eric~F Lock, and Dennis~F
  Lock.
\newblock {\em Statistics: Unlocking the power of data}.
\newblock John Wiley \& Sons, 2020.

\bibitem{dino}
Brandon Lucia and Benjamin Ransford.
\newblock A simpler, safer programming and execution model for intermittent
  systems.
\newblock In {\em the 36th annual ACM SIGPLAN conference on Programming
  Language Design and Implementation (PLDI'15)}, pages 575--585, 2015.

\bibitem{luckow2014exact}
Kasper Luckow, Corina~S P{\u{a}}s{\u{a}}reanu, Matthew~B Dwyer, Antonio
  Filieri, and Willem Visser.
\newblock Exact and approximate probabilistic symbolic execution for
  nondeterministic programs.
\newblock In {\em the 29th ACM/IEEE international conference on Automated
  software engineering (ASE'14)}, pages 575--586, 2014.

\bibitem{alpaca}
Kiwan Maeng, Alexei Colin, and Brandon Lucia.
\newblock Alpaca: intermittent execution without checkpoints.
\newblock In {\em the 32nd annual ACM Conference on Object-Oriented
  Programming, Systems, Languages and Applications (OOPSLA'17)}, pages 1--30,
  2017.

\bibitem{chinchilla}
Kiwan Maeng and Brandon Lucia.
\newblock Adaptive dynamic checkpointing for safe efficient intermittent
  computing.
\newblock In {\em 13th $\{$USENIX$\}$ Symposium on Operating Systems Design and
  Implementation ($\{$OSDI$\}$ 18)}, pages 129--144, 2018.

\bibitem{catnap}
Kiwan Maeng and Brandon Lucia.
\newblock Adaptive low-overhead scheduling for periodic and reactive
  intermittent execution.
\newblock In {\em the 41st ACM SIGPLAN Conference on Programming Language
  Design and Implementation (PLDI'20)}, pages 1005--1021, 2020.

\bibitem{sceptic}
Andrea Maioli, Luca Mottola, Muhammad~Hamad Alizai, and Junaid~Haroon Siddiqui.
\newblock On intermittence bugs in the battery-less internet of things (wip
  paper).
\newblock In {\em the 20th ACM SIGPLAN/SIGBED International Conference on
  Languages, Compilers, and Tools for Embedded Systems (LCTES'19)}, pages
  203--207, 2019.

\bibitem{coala}
Amjad~Yousef Majid, Carlo~Delle Donne, Kiwan Maeng, Alexei Colin, Kasim~Sinan
  Yildirim, Brandon Lucia, and Przemys{\l}aw Pawe{\l}czak.
\newblock Dynamic task-based intermittent execution for energy-harvesting
  devices.
\newblock {\em ACM Transactions on Sensor Networks (TOSN)}, 16(1):1--24, 2020.

\bibitem{mishra2015charging}
Deepak Mishra, Swades De, and Kaushik~R Chowdhury.
\newblock Charging time characterization for wireless rf energy transfer.
\newblock {\em IEEE Transactions on Circuits and Systems II: Express Briefs},
  62(4):362--366, 2015.

\bibitem{naderi2015wireless}
M~Yousof Naderi, Kaushik~R Chowdhury, and Stefano Basagni.
\newblock Wireless sensor networks with rf energy harvesting: Energy models and
  analysis.
\newblock In {\em 2015 IEEE Wireless Communications and Networking Conference
  (WCNC)}, pages 1494--1499. IEEE, 2015.

\bibitem{wispcam}
Saman Naderiparizi, Aaron~N Parks, Zerina Kapetanovic, Benjamin Ransford, and
  Joshua~R Smith.
\newblock Wispcam: A battery-free rfid camera.
\newblock In {\em IEEE International Conference on RFID (RFID'15)}, pages
  166--173. IEEE, 2015.

\bibitem{camaroptera}
Matteo Nardello, Harsh Desai, Davide Brunelli, and Brandon Lucia.
\newblock Camaroptera: A batteryless long-range remote visual sensing system.
\newblock In {\em the 7th International Workshop on Energy Harvesting and
  Energy-Neutral Sensing Systems (ENSsys'19)}, pages 8--14, 2019.

\bibitem{puasuareanu2010symbolic}
Corina~S P{\u{a}}s{\u{a}}reanu and Neha Rungta.
\newblock Symbolic pathfinder: symbolic execution of java bytecode.
\newblock In {\em the 25th IEEE/ACM international conference on Automated
  software engineering (ASE'10)}, pages 179--180, 2010.

\bibitem{powercast}
{Powercast Corp.}
\newblock Powercast hardware.
\newblock \url{http://www.powercastco.com}, 2021.
\newblock Last accessed: Aug. 10, 2021.

\bibitem{P2110-EVB}
{Powercast Corp.}
\newblock Powercast hardware.
\newblock
  \url{https://www.powercastco.com/wp-content/uploads/2016/11/p2110-evb1.pdf},
  2021.
\newblock Last accessed: Aug. 10, 2021.

\bibitem{R}
{R Core Team}.
\newblock {\em R: A Language and Environment for Statistical Computing}.
\newblock R Foundation for Statistical Computing, Vienna, Austria, 2017.

\bibitem{mementos}
Benjamin Ransford, Jacob Sorber, and Kevin Fu.
\newblock Mementos: System support for long-running computation on {RFID-scale}
  devices.
\newblock In {\em the 16th international conference on Architectural support
  for programming languages and operating systems (ASPLOS'11)}, pages 159--170,
  2011.

\bibitem{enhants}
{RAWDAD}.
\newblock The columbia/enhants dataset.
\newblock \url{https://crawdad.org/columbia/enhants/20110407/}, 2022.
\newblock Last accessed: 2022.

\bibitem{coati}
Emily Ruppel and Brandon Lucia.
\newblock Transactional concurrency control for intermittent, energy-harvesting
  computing systems.
\newblock In {\em the 40th ACM SIGPLAN Conference on Programming Language
  Design and Implementation (PLDI'19)}, pages 1085--1100, 2019.

\bibitem{Saleae}
{Saleae}.
\newblock Saleae logic pro 16 analyzer.
\newblock
  \url{https://support.saleae.com/datasheets-and-specifications/datasheets},
  2021.
\newblock Last accessed: Aug. 10, 2021.

\bibitem{wisp}
Alanson~P Sample, Daniel~J Yeager, Pauline~S Powledge, Alexander~V Mamishev,
  and Joshua~R Smith.
\newblock Design of an rfid-based battery-free programmable sensing platform.
\newblock {\em IEEE Transactions on Instrumentation and Measurement},
  57(11):2608--2615, 2008.

\bibitem{ibis}
Milijana Surbatovich, Limin Jia, and Brandon Lucia.
\newblock I/o dependent idempotence bugs in intermittent systems.
\newblock In {\em the 34th annual ACM Conference on Object-oriented
  Programming, Systems, Languages and Applications (OOPSLA'19)}, pages 1--31,
  2019.

\bibitem{ocelot}
Milijana Surbatovich, Limin Jia, and Brandon Lucia.
\newblock Automatically enforcing fresh and consistent inputs in intermittent
  systems.
\newblock In {\em Proceedings of the 42nd ACM SIGPLAN International Conference
  on Programming Language Design and Implementation}, pages 851--866, 2021.

\bibitem{EnergyTrace}
{T}exas {I}nstruments.
\newblock {EnergyTrace Technology}.
\newblock \url{https://www.ti.com/tool/energytrace}, 2021.

\bibitem{msp430datasheet}
{T}exas {I}nstruments.
\newblock {MSP} datasheet.
\newblock
  \url{https://www.ti.com/sc/docs/products/micro/msp430/userguid/as_5.pdf},
  2021.

\bibitem{MSP430FR5994}
{T}exas {I}nstruments.
\newblock {MSP430FR5994 Mixed-Signal Microcontroller}.
\newblock \url{https://www.ti.com/product/MSP430FR5994}, 2021.

\bibitem{FRAM}
{Texas Instruments, Inc.}
\newblock {FRAM} faqs.
\newblock \url{http://www.ti.com/lit/ml/slat151/slat151.pdf}, 2021.
\newblock Last accessed: 2021.

\bibitem{capband}
Hoang Truong, Shuo Zhang, Ufuk Muncuk, Phuc Nguyen, Nam Bui, Anh Nguyen, Qin
  Lv, Kaushik Chowdhury, Thang Dinh, and Tam Vu.
\newblock Capband: Battery-free successive capacitance sensing wristband for
  hand gesture recognition.
\newblock In {\em the 16th ACM Conference on Embedded Networked Sensor Systems
  (SenSys'18)}, pages 54--67, 2018.

\bibitem{ratchet}
Joel Van Der~Woude and Matthew Hicks.
\newblock Intermittent computation without hardware support or programmer
  intervention.
\newblock In {\em the 12th $\{$USENIX$\}$ Symposium on Operating Systems Design
  and Implementation ($\{$OSDI$\}$ 16)}, pages 17--32, 2016.

\bibitem{van1987simulated}
Peter~JM Van~Laarhoven and Emile~HL Aarts.
\newblock Simulated annealing.
\newblock In {\em Simulated annealing: Theory and applications}, pages 7--15.
  Springer, 1987.

\bibitem{ink}
Kas{\i}m~Sinan Y{\i}ld{\i}r{\i}m, Amjad~Yousef Majid, Dimitris Patoukas, Koen
  Schaper, Przemyslaw Pawelczak, and Josiah Hester.
\newblock {InK}: Reactive kernel for tiny batteryless sensors.
\newblock In {\em the 16th ACM Conference on Embedded Networked Sensor Systems
  (SenSys'18)}, pages 41--53, 2018.

\end{thebibliography}


\end{document}